\documentclass[a4paper,11pt]{article}
\pdfoutput=1

\usepackage{amsmath}
\usepackage{jheppub}

\usepackage[utf8]{inputenc}
\usepackage[T1]{fontenc}
\usepackage[kerning,spacing]{microtype}

\usepackage{graphicx}

\usepackage{amsbsy,amssymb}
\usepackage{mathtools}
\usepackage{xspace}

\usepackage{dcolumn}
\usepackage{bm}
\usepackage{slashed}
\usepackage{color}
\usepackage[font=small]{caption}
\usepackage{subcaption}
\usepackage{todonotes}
\presetkeys{todonotes}{inline}{}

\usepackage{booktabs}

\newcommand{\Higgs}{\ensuremath{\mathrm{H}}\xspace}

\newcommand{\Hj}{\Higgs\!+\! 1\! jet\xspace}

\newcommand{\Hjj}{\Higgs\!+\! 2\! jets\xspace}
\newcommand{\Hjjj}{\Higgs\!+\! 3\! jets\xspace}

\newcommand{\Hjets}{\Higgs\!+\! 1,\,2 and 3\! jets\xspace}
\newcommand{\Hnj}{\Higgs\!+\! $n$\! jets\xspace}

\newcommand{\hthatprime}{\hat{H}^\prime_T}

\newcommand{\GeV}{\ensuremath{\mathrm{GeV}}\xspace}
\newcommand{\TeV}{\ensuremath{\mathrm{TeV}}\xspace}

\title{Full mass dependence in Higgs boson production in association
  with jets at the LHC and FCC}
\author[a]{Nicolas Greiner,}
\author[b]{Stefan H\"oche,}
\author[c]{Gionata Luisoni,}
\author[a]{Marek Sch\"onherr,}
\author[d]{Jan-Christopher Winter\hphantom{,}}
\affiliation[a]{Physik-Institut, Universit\"at Z\"urich, Wintherturerstrasse 190, CH-8057 Z\"urich, Switzerland}
\affiliation[b]{SLAC National Accelerator Laboratory, Menlo Park, CA 94025, USA}
\affiliation[c]{Theoretical Physics Department, CERN, Geneva, Switzerland}
\affiliation[d]{Department of Physics and Astronomy, Michigan State University, East Lansing, MI 48824, USA}

\preprint{{\small%
  CERN-TH-2016-171\\
  \hphantom{.}\hfill MSUHEP-160727\\
  \hphantom{.}\hfill SLAC-PUB-16778\\
  \hphantom{.}\hfill ZH-TH-28/16\\
  \hphantom{.}\hfill MCNET-16-32}
}


\abstract{The first computation of Higgs production in association
  with three jets at NLO in QCD has recently been performed using the
  effective theory, where the top quark is treated as an infinitely
  heavy particle and integrated out. This approach is restricted to
  the regions in phase space where the typical scales are not larger
  than the top quark mass. Here we investigate this statement at a
  quantitative level by calculating the leading-order contributions to
  the production of a Standard Model Higgs boson in association with
  up to three jets taking full top-quark and bottom-quark mass
  dependence into account. We find that the transverse momentum of the
  hardest particle or jet plays a key role in the breakdown of the
  effective theory predictions, and that discrepancies can easily
  reach an order of magnitude for transverse momenta of about 1 TeV.
  The impact of bottom-quark loops are found to be visible in the
  small transverse momentum region, leading to corrections of up to 5
  percent. We further study the impact of mass corrections when VBF
  selection cuts are applied and when the center-of-mass energy is
  increased to 100 TeV.}

\begin{document}

\maketitle

\section{Introduction}\label{sec:intro}

The gluon fusion mechanism yields the largest contribution to the
production cross section of a Standard Model (SM) Higgs boson. However, the
fact that already at leading order (LO) this process is mediated by a
closed loop of heavy fermions, in other words it is a loop-induced
process, leads to a tremendous complication in the computation of
theoretical predictions and higher order corrections. This holds not
only for the production of a Higgs boson alone, but also and
especially for the calculation of its production in association with
jets.

When the mass of the fermions is much larger than the Higgs boson
mass, the heavy fermion can be integrated out and the coupling between
gluons and the Higgs can be described by an effective vertex
\cite{Wilczek:1977zn}, simplifying the calculations
considerably. Since the top quark is giving the dominant contribution
in the heavy fermion loops, this approximation is also referred to as
the infinite top-quark mass limit. The validity of the effective
theory is however limited. In particular it breaks down when the
momentum flow through effective vertex becomes of the order as the
fermion masses.

This behaviour can be understood better by comparing the high-energy
limit of a pointlike gluon-gluon Higgs interaction, with a resolved
interaction mediated via a loop. The latter provides a form factor
responsible for softening the amplitude in this limit. More
specifically one has to consider the transverse momentum behaviour of
the amplitude for producing a Higgs boson out of two off-shell gluons
in the effective and in the full theory
~\cite{Catani:1990eg,Hautmann:2002tu,Pasechnik:2006du,Marzani:2008az}. The
contribution in the limit of large gluon transverse momenta (much
larger than the heavy-quark mass) is suppressed by the massive quark
loop in the full theory, whereas in a pointlike interaction the same
transverse momenta are allowed to reach the kinematic limit given by
the center-of-mass energy $\sqrt{s}$. In the high energy, limit this leads to
a different scaling for the two predictions in terms of the leading
logarithmic contribution in $(m_H^2/s)$, where $m_H$ is the
Higgs boson mass. The effective theory has a double logarithmic
scaling, whereas the full theory scales as a single logarithm of
$(m_H^2/s)$. Recently the corresponding scaling in terms of the
Higgs boson transverse momentum $p_{T,\,H}$ was derived in the same
high energy limit~\cite{Forte:2015gve,Caola:2016upw}, finding that, as
$p_{T,\,H}\to\infty$, the distribution drops as
$(p_{T,\,H}^2)^{-1}$ in the effective theory, whereas it goes as
$(p_{T,\,H}^2)^{-2}$ in the full theory.

The predictions obtained in the infinite top-quark mass limit are
therefore an increasingly poor approximation as the transverse
momentum of the Higgs boson increases and becomes larger than roughly
the top quark mass. This affects an increasing fraction of the phase
space when the Higgs boson is produced in association with several
jets. The case in which the Higgs boson is produced via the gluon fusion
process in association with at least two further jets represents also
the most relevant irreducible background to Higgs boson production via
vector boson fusion (VBF). These two production mechanisms can be
distinguished introducing topological cuts, which may however enhance
even more the portion of phase space in which the effective
gluon-gluon-Higgs theory is a poor approximation of the full theory
prediction.

The purpose of this paper is to investigate the range of validity and the
breakdown of the effective theory approach at a more quantitative level,
when the Higgs boson is produced in association with up to three
jets. In other words, we will pursue the question, if and to what
extent the large top-quark mass limit is justified for higher jet
multiplicities. Taking full top- and bottom-quark mass dependence in
the loop into account, we perform a leading order calculation for the
production of a Higgs boson in association with up to three jets and
compare this to predictions from the effective theory.

A precise treatment of massive bottom quarks at leading order requires
the use of four flavor PDFs and the corresponding removal of initial
state bottom quarks. Furthermore, massive b-quarks also invoke a
Higgs-bottom Yukawa coupling, which leads to tree-level contributions
with massive bottom quarks in the final state. However, in this paper
we are interested in the mass effects caused by massive quarks in the
loops. In other words we want to determine the effect of taking bottom
quarks into account, compared to predictions in which only top quarks
are considered. Therefore we keep the external quarks massless and
leave the aforementioned approach for further studies.

Leading order results in the full theory for up to two jets have been
known for some
time~\cite{Georgi:1977gs,Baur:1989cm,Djouadi:1991tka,Spira:1995rr,
  DelDuca:2001eu,DelDuca:2001fn}, and partial results for Higgs boson
plus three jet production were first computed
in~\cite{Campanario:2013mga}, whereas lately multijet merged
predictions for up to two or three jets with full mass dependence
were computed in~\cite{Buschmann:2014sia} and~\cite{Frederix:2016cnl}
respectively. As already anticipated, very recently predictions of
mass effects beyond LO on the Higgs boson transverse momentum spectrum
became available too~\cite{Caola:2016upw}.  A different study
investigated instead the effect of light-quark mediated
contributions~\cite{Melnikov:2016emg}.

Studying the effects of mass corrections to the infinite top-quark mass
limit becomes even more important for proton colliders with very large
center-of-mass energies. Recently, an analysis similar to the one we
present here was performed in the context of a comprehensive report
about physics at the Future Circular Collider (FCC) for a
center-of-mass energy of 100~\TeV~\cite{Contino:2016spe}.

The paper is structured in the following way: in
Section~\ref{sec:calculational_setup} we present the setup used to
perform the computation, the choice of the phenomenological parameters
and the cuts we applied. Section~\ref{sec:total_xsec} is dedicated to
the total cross section results for LHC and FCC, whereas the results
at the differential level are presented in
Section~\ref{sec:differential_dist}.  In Section~\ref{sec:conclusions}
we conclude and offer an outlook on possible future improvements.

\section{Calculational setup}
\label{sec:calculational_setup}

In this paper we will compare predictions for the production of a
Higgs boson in association with one, two or three jets at LO and
next-to-leading order (NLO) in the effective Higgs-gluon theory,
already computed in~\cite{Cullen:2013saa,Greiner:2015jha}, with predictions at LO in
the full SM for all three multiplicities. The latter were computed in
two different manners: once considering only massive top-quark loop
contributions, and once taking into account both massive top- and
bottom-quarks running in the loop.

For \Hj there are two different partonic channels which have to be
considered:
\begin{flalign}
 q\,\bar{q} & \to H\,g\,,\nonumber\\
 g\,g       & \to H\,g\;.
\end{flalign}
For both \Hjj and \Hjjj there are instead four different independent
subprocesses, namely
\begin{flalign}
 q\,\bar{q} & \to H\,q'\,\bar{q}'\,(g)\,, \nonumber \\
 q\,\bar{q} & \to H\,q \,\bar{q} \,(g)\,, \nonumber \\
 q\,\bar{q} & \to H\,g \,g\,(g)\,,        \nonumber \\
 g\,g       & \to H\,g \,g\,(g)\;.
\end{flalign}
All the remaining subprocesses are related by crossing symmetry.

Both the one-loop amplitudes for the NLO effective theory results as
well as the one-loop amplitudes for the LO results with mass
dependence were generated using {\sc{GoSam}}
\cite{Cullen:2011ac,Cullen:2014yla}, a publicly available package for
the automated generation of one-loop amplitudes.  It is based on an
algebraic generation of d-dimensional integrands using a Feynman
diagrammatic approach, employing {\sc{QGRAF}}~\cite{Nogueira:1991ex}
and {\sc{FORM}}~\cite{Vermaseren:2000nd,Kuipers:2012rf} for the
diagram generation, and {\sc{Spinney}}~\cite{Cullen:2010jv},
{\sc{Haggies}}~\cite{Reiter:2009ts} and {\sc{FORM}} to write an
optimized Fortran output. For the reduction of the tensor integrals we
use
{\sc{Ninja}}~\cite{Mastrolia:2012bu,Peraro:2014cba,vanDeurzen:2013saa},
a tool for the integrand reduction via Laurent expansion.
Alternatively one can use other reduction techniques such as integrand
reduction using the OPP
method~\cite{Ossola:2006us,Mastrolia:2008jb,Ossola:2008xq} as
implemented in {\sc{Samurai}}~\cite{Mastrolia:2010nb} or using methods
of tensor reduction as offered by the
{\sc{Golem95}}~\cite{Heinrich:2010ax,Binoth:2008uq,Cullen:2011kv,Guillet:2013msa}
library. The remaining scalar integrals have been evaluated using
{\sc{OneLoop}}~\cite{vanHameren:2010cp}.

For the NLO prediction in the effective theory the tree-level
amplitudes for the Born and real radiation contribution, the
subtraction terms and their integrated counterpart were computed with
\textsc{Sherpa} \cite{Gleisberg:2008ta} and the matrix element
generator \textsc{Comix}~\cite{Gleisberg:2008fv,Hoeche:2014xx}.
\textsc{Sherpa} and {\sc{GoSam}} were linked via the Binoth Les
Houches Accord interface~\cite{Binoth:2010xt,Alioli:2013nda}.

Because of the high statistics needed for such a large multiplicity
final state, the Monte Carlo events are stored in the form of
\textsc{Root} Ntuples. They are generated by \textsc{Sherpa} and were
first used in the context of vector boson production in association
with jets~\cite{Bern:2013zja}. Very recently first studies appeared
about possible extensions for NNLO
computations~\cite{Badger:2016bpw,Heinrich:2016jad}.

For the calculation in the effective theory, sets of Ntuples files
with Born (B), virtual (V), integrate subtraction term (I) and real
minus subtraction term (RS) type of events have been generated for
\Hjets at the center-of-mass energies of 13 and 100
TeV~\footnote{Similar sets of Ntuples were also generated at 8 and
  14~\TeV.}. The events were generated such that jets can be clustered
using the $k_T$ or anti-$k_T$
algorithm~\cite{Cacciari:2005hq,Cacciari:2008gp} as implemented in the
\textsc{FastJet} package~\cite{Cacciari:2011ma} and with radii that
can vary between $R=0.1$ and $R=1$. At 13~\TeV a minimal generation
cut was imposed on the jets by requiring
\begin{equation}
  \label{gencuts:basic}
  p_{T,\,\mathrm{jet}}\;>\;25\mathrm{\:GeV}\qquad\mbox{and}\qquad
  |\eta_\mathrm{jet}|\;<	\;4.5~\,.
\end{equation}
Because of the much wider rapidity span available, at 100~\TeV the
generation cuts are:
\begin{equation}
  \label{gencuts:basic}
  p_{T,\,\mathrm{jet}}\;>\;25\mathrm{\:GeV}\qquad\mbox{and}\qquad
  |\eta_\mathrm{jet}|\;<\;10~\,,
\end{equation}
which allows to post-process the events in every analysis with more
exclusive cuts.  Furthermore they allow the user to change a
posteriori both the renormalization and the factorization scales as
well as the choice of the parton distribution functions
(PDFs) \footnote{All the set of Ntuples at 8, 13, 14 and 100~\TeV are
  publicly available on EOS via the following link:
  https://eospublic.cern.ch/eos/theory/project/GoSam/}.  More details
about the format of the Ntuples and an extension of their content used
for this work is discussed in Appendix~\ref{app:ntuples}.

\subsection{Generation of Ntuples incorporating finite-mass effects}

The set of Ntuples with the full mass dependence were generated
starting from the available Born type Ntuples used for the study
presented in~\cite{Greiner:2015jha}. These Ntuples contain the Born
matrix element weight in the effective theory. To obtain a set of LO
Ntuples in the full theory we have therefore re-weighted these events
using the matrix elements with the full quark mass dependence.

Since the effective theory is obtained by assuming an infinitely heavy
top quark, which is integrated out, the one-loop amplitudes could be
checked in a robust way by setting the top mass to large values and
observing that the effective theory result is reproduced.  We have
checked this behavior numerically by setting the top mass to 10 TeV
and found an agreement at the sub-permille level between the one-loop
amplitudes of the full theory with the tree-level amplitudes of the
effective theory for random phase space points.  This is a strong
consistency check for the whole setup, in particular of course for the
correctness of the one-loop amplitudes.

\subsection{Physical parameters}

In the following we will present numerical results for center-of-mass
energies of 13~\TeV and for a possible future collider at 100~\TeV.  As
input parameters we use\\
\begin{align}
  \centering
  m_H &= 125.0~\GeV,   \quad &   m_t &= 172.3~\GeV\,,\nonumber\\
  m_Z &= 91.1876~\GeV, \quad &     v &= 246~\GeV\,,  \\
  \alpha_\textsc{qed} &= 1/128.8022\,.\nonumber
\end{align}
When we consider also bottom quark loops, the bottom-quark mass was
set to $m_b=4.75$~\GeV in the propagator mass and to $m_b(m_H) =
3.38$~\GeV in the Yukawa coupling \cite{Hamilton:2015nsa}. This allows
us to quantify the effect due to bottom-quark loops and its
interference with top-quark loops. For external partons we keep
working with $n_f=5$ light active flavours.

From the input parameters listed above we derive the corresponding
values for $m_W$ and $\sin_w$ that enter in the definition of the
gluon-Higgs coupling in the effective theory.  We define our central
renormalization and factorization scale to be
\begin{equation}
\mu_{F}=\mu_{R}\;\equiv\;\frac{\hthatprime}{2}\;=\;
\frac{1}{2}\left(\sqrt{m_H^2+p_{T,\,H}^2}+\sum_{i}|p_{T,i}|\right)\;,
\end{equation}
where the sum is understood to run over partons rather than over
jets. Scale uncertainties are obtained by varying both scales
simultaneously by factors of $0.5$ and $2$ around the central
value. The strong coupling constant is calculated at this scale and
taken according to the CT14NLO pdf set~\cite{Dulat:2015mca}.

We investigate two different set of cuts, one that is suited for a
general analysis of the gluon fusion scenario with only a basic set of
cuts to render the cross section finite, and a second set, which is
more suitable in the context of the vector boson fusion scenario.  The
baseline cuts for the jets consists of
\begin{equation}
 p_{T,\,\mathrm{jet}}\;>\;30~\GeV~,\qquad |y_\mathrm{jet}|\;<\;4.4~.
 \label{cuts:basic}
\end{equation}
In addition to these cuts, to investigate the vector boson fusion
(VBF) scenario, we further demand
\begin{equation}
  m_{j_1 j_2}\;>\;400~\GeV~,\qquad \left|\Delta y_{j_1,\,j_2}\right|\;>\;2.8~,
  \label{cuts:vbf}
\end{equation}
where $j_1$ and $j_2$ are the leading jets for a given tagging
scheme. We will refer to them as \textit{tagging jets} in the
following. In the next sections, we will mainly consider a $p_T$
jet-tagging strategy, in which jets are order by decreasing transverse
momentum. In this case $j_1$ and $j_2$ are the leading and the
second-leading transverse momentum jets. For some specific observables,
we will however also consider a $y$ jet-tagging scheme in which the
tagging jets are defined as the two jets with the most forward and
most backward rapidity.

\section{Total cross sections including top and bottom quark contributions}
\label{sec:total_xsec}

\begin{table}[t!]
\centering\small
\begin{tabular}{|l||c|c|c|}
  \hline
  Numbers in [pb]
  & \multicolumn{2}{c|}{$p_{T,\,\mathrm{jet}}>30~\GeV$\phantom{\Big|}}
  & $p_{T,\,\mathrm{jet}}>100~\GeV$ \\
  \hline
  $\sqrt{s}$
  & $13\, \TeV$
  & $100\,\TeV$
  & $100\,\TeV$
  \phantom{\Big|}\\
  \hline
  \multicolumn{4}{|c|}{$\mathrm{H}\!+\!1$ jet\phantom{\Big|}}\\
  \hline
  $\sigma_{\mathrm{LO,\,eff.}}$    \phantom{\Big|} & $8.06^{+38\%}_{-26\%}$ & $196^{+21\%}_{-17\%}$ & $55.7^{+24\%}_{-19\%}$ \\
  $\sigma_{\mathrm{NLO,\,eff.}}$   \phantom{\Big|} & $13.3^{+15\%}_{-15\%}$ & $315^{+11\%}_{-10\%}$ & $88.8^{+11\%}_{-11\%}$ \\
  $\sigma_{\mathrm{LO},\,m_{t,b}}$  \phantom{\Big|} & $8.35^{+38\%}_{-26\%}$ & $200^{+20\%}_{-17\%}$ & $52.3^{+24\%}_{-19\%}$ \\
  $\sigma_{\mathrm{LO},\,m_{t}}$    \phantom{\Big|} & $8.40^{+38\%}_{-26\%}$ & $201^{+20\%}_{-17\%}$ & $51.3^{+24\%}_{-18\%}$ \\
  \hline
  \multicolumn{4}{|c|}{$\mathrm{H}\!+\!2$ jets\phantom{\Big|}}\\
  \hline
  $\sigma_{\mathrm{LO,\,eff.}}$    \phantom{\Big|} & $2.99^{+58\%}_{-34\%}$ & $124^{+39\%}_{-27\%}$ & $16.5^{+41\%}_{-28\%}$ \\
  $\sigma_{\mathrm{NLO,\,eff.}}$   \phantom{\Big|} & $4.55^{+13\%}_{-18\%}$ & $156^{+3\%}_{-10\%}$  & $23.3^{+9\%}_{-13\%}$  \\
  $\sigma_{\mathrm{LO},\,m_{t,b}}$  \phantom{\Big|} & $3.08^{+58\%}_{-34\%}$ & $121^{+39\%}_{-26\%}$ & $13.2^{+41\%}_{-27\%}$ \\
  $\sigma_{\mathrm{LO},\,m_{t}}$    \phantom{\Big|} & $3.05^{+58\%}_{-34\%}$ & $120^{+39\%}_{-26\%}$ & $13.0^{+41\%}_{-27\%}$ \\
  \hline
  \multicolumn{4}{|c|}{$\mathrm{H}\!+\!3$ jets\phantom{\Big|}}\\
  \hline
  $\sigma_{\mathrm{LO,\,eff.}}$   \phantom{\Big|} & $0.98^{+76\%}_{-41\%}$ & $70.4^{+56\%}_{-34\%}$ & $5.13^{+56\%}_{-34\%}$ \\
  $\sigma_{\mathrm{NLO,\,eff.}}$  \phantom{\Big|} & $1.45^{+11\%}_{-22\%}$ & $72.0^{-16\%}_{-7\%}$  & $6.52^{+2\%}_{-14\%}$  \\
  $\sigma_{\mathrm{LO},\,m_{t,b}}$ \phantom{\Big|} & $1.00^{+77\%}_{-41\%}$ & $63.3^{+56\%}_{-34\%}$ & $3.38^{+57\%}_{-34\%}$ \\
  $\sigma_{\mathrm{LO},\,m_{t}}$   \phantom{\Big|} & $0.99^{+77\%}_{-41\%}$ & $62.7^{+56\%}_{-34\%}$ & $3.32^{+56\%}_{-34\%}$ \\
  \hline
\end{tabular}
\caption{\label{tab:xsec_ggf}%
  Total inclusive cross sections for the production of a Higgs boson
  in association with one, two or three jets at LO and NLO in QCD in
  the effective theory and at LO in the full SM for massive top- and
  bottom-quarks and for massive top quarks only. Numbers are reported
  for center-of-mass energies of 13 and 100~\TeV and 2 choices of
  transverse momentum cuts on the jets, namely
  $p_{T,\,\mathrm{jet}}>30$ and $100~\GeV$. The uncertainty estimates
  are obtained from standard scale variations.}
\end{table}

We start the discussion of the numerical results with a comparison of
the total cross sections, for which we consider the effective theory
predictions at LO and NLO (labeled as $\sigma_{\mathrm{LO,\,eff}}$
and $\sigma_{\mathrm{NLO,\,eff}}$ respectively) and compare them with
the full theory results at leading order when considering both
top-quark and bottom-quark loops, called
$\sigma_{\mathrm{LO},\,m_{t,b}}$, as well as with the case where only
top-quark loops are taken into account, labeled
$\sigma_{\mathrm{LO},\,m_{t}}$. In Table~\ref{tab:xsec_ggf} we
summarize the results for the different jet multiplicities and for
center-of-mass energies of 13~\TeV and 100~\TeV. The effect of varying
the renormalization and factorization scales by factors of $0.5$ and
$2$ is reported as a relative variation with respect to the nominal
value. As expected all the LO results both in the effective and in the
full theory suffer from large scale dependencies, which become as
large as $70\%$ for \Hjjj, and which reduce to $10-20\%$ at NLO in the
effective theory.

\begin{figure}[t!]
  \centering
  \includegraphics[width=0.49\textwidth]{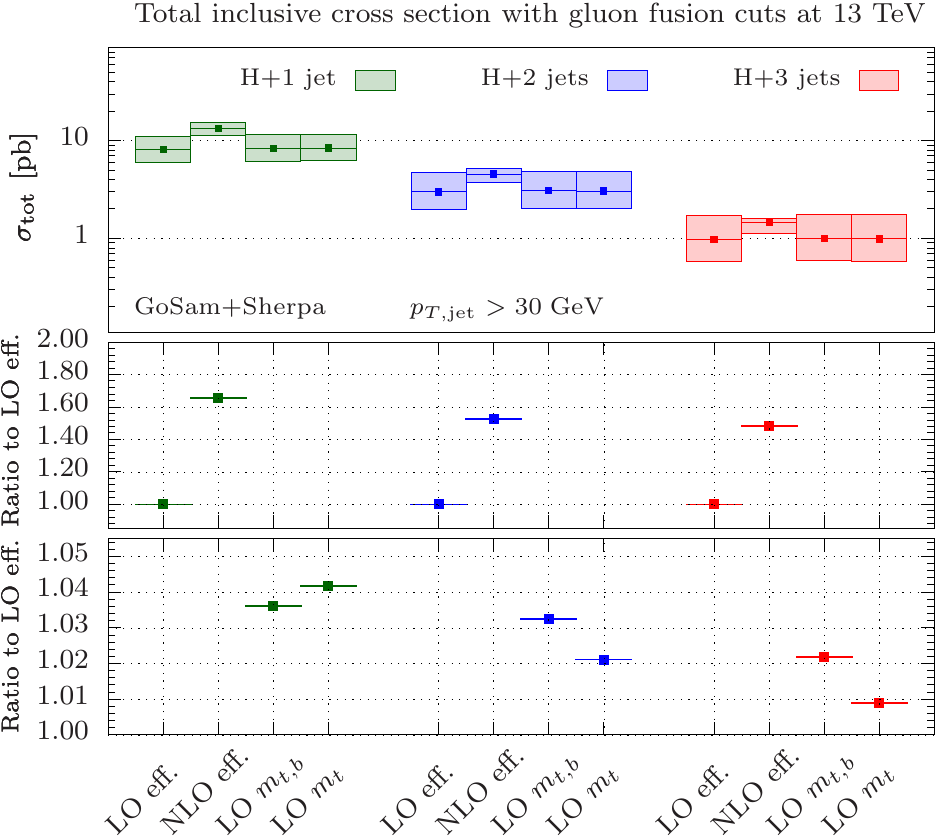}
  \hfill
  \includegraphics[width=0.49\textwidth]{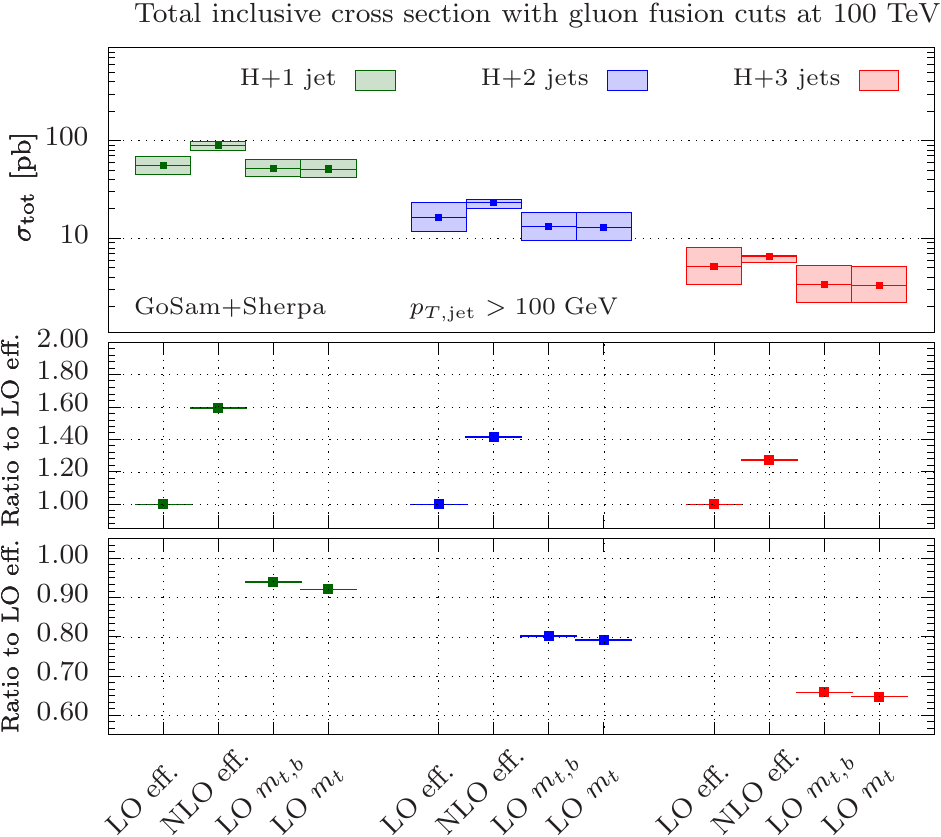}
  \caption{\label{fig:inclxs}%
    Inclusive cross sections for \Hj, \Hjj and \Hjjj production at
    center-of-mass energies of $13~\TeV$ and $100~\TeV$ shown to the
    left and right, respectively. The width of the bands shows the
    associated scale uncertainty.}
\end{figure}

The results of Table~\ref{tab:xsec_ggf} are visualized in
Figure~\ref{fig:inclxs}, where we also include the ratios to the
leading order result in the effective theory. For a better visibility
we show two different ratio plots, both normalized to the LO result in
the effective theory. The upper ratio shows the $K$-factor between LO
and NLO in the effective theory. The lower one, with a much smaller
range on the y-axis, highlights the differences between the LO in the
effective and in the full theory.

By combining the LO prediction that includes the exact top-quark mass
dependence and the NLO $K$-factor from the effective theory
calculation, one could estimate the Higgs boson plus multi-jet cross
section with exact top-quark mass dependence at NLO.  This approach
was used successfully for lower-multiplicity calculations
in~\cite{Buschmann:2014sia}. The much more demanding computation of
the exact mass dependence of the cross section has been performed for
inclusive Higgs production at NLO~\cite{Spira:1995rr} and even at
NNLO~\cite{Harlander:2009bw,Pak:2009bx,Harlander:2009mq,
  Pak:2009dg,Harlander:2009my,Harlander:2012hf}, but the exact result
at NLO is not within reach for the large jet multiplicities considered
here.  A combination of the LO result with full top-quark mass
dependence and the NLO $K$-factor from the effective theory would
constitute the current best estimate of Higgs boson production in
association with up to three jets.  However, we refrain from quoting
the corresponding number, as it does not add new information to our
existing predictions.

Focusing on the central values, we observe that the leading-order
contribution in the effective theory agrees in general very well with
the predictions based on the full theory. Taking bottom-quark loops into
account leads to corrections, which are as small as one percent for all
three final-state multiplicities we are considering, and, as expected,
they become even smaller at $100~\TeV$. However, it is interesting to
note the change in the sign of these corrections with increasing jet
multiplicity. While for \Hj production at $13~\TeV$ the cross section is
reduced when bottom-quark loop contributions are included, for \Hjj
and \Hjjj the cross section increases instead. This is clearly visible
in the first column of Table~\ref{tab:xsec_ggf} and displayed in the
second ratio plot on the left in Figure~\ref{fig:inclxs}. If we
compare the predictions at $100~\TeV$ given in
Table~\ref{tab:xsec_ggf} for the two different transverse
momentum cuts applied to the jets, we observe a sign flip in the
interference effects for \Hj. While for $p_{T,\,\mathrm{jet}}>30~\GeV$, the
pattern is similar to the $13~\TeV$ results, increasing the minimum
$p_T$ of the jet instead yields a small positive overall
contribution once both quark-loop contributions of the heaviest
generation are included. This means that the effect of the
interference on the cross section is destructive in the low transverse
momentum region, whereas at high $p_T$ the sum of top-quark and
bottom-quark contributions leads to constructive interference
effects. We will come back to this in Section~\ref{sec:bottom_vs_top},
where we discuss the impact of these interference effects on
differential distributions.

\section{Heavy-quark mass effects in differential distributions}
\label{sec:differential_dist}

Recalling the 100~\TeV collider results from Table~\ref{tab:xsec_ggf},
we have already seen that an increase of the jet $p_T$ threshold
leads to a noticeable change of the total Higgs boson plus jet cross
sections. By studying differential cross sections for various classes
of observables, we want to identify the phase-space regions that
receive important corrections as a result of the finite-mass treatment
of the heavy-quark loops. For a broader understanding of this issue,
we consider different scenarios that are of relevance to ongoing and
future hadron collider experiments.

\subsection{LHC predictions for 13 TeV collisions}

\begin{figure}[th!]
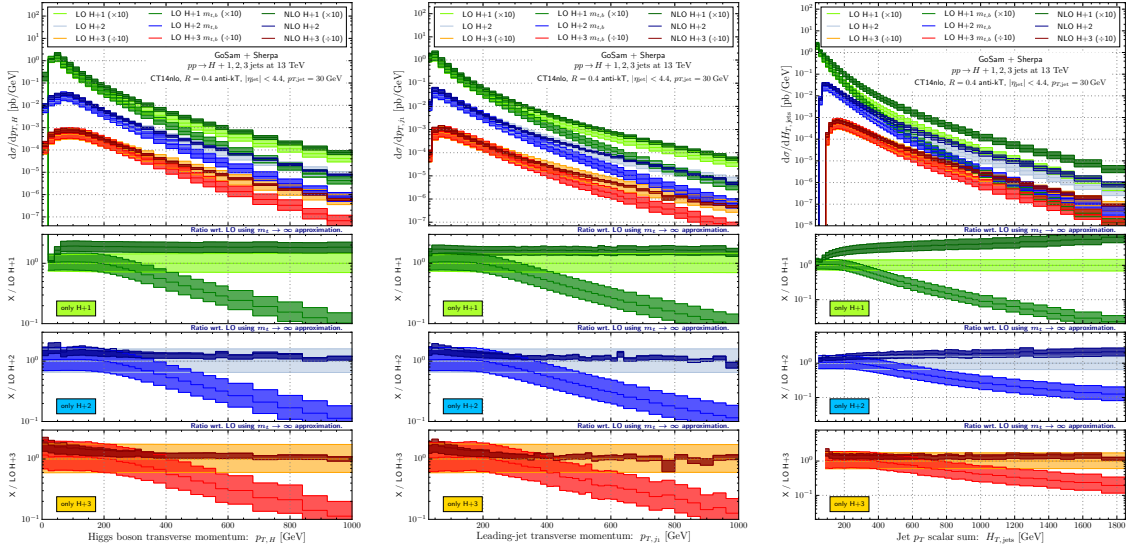
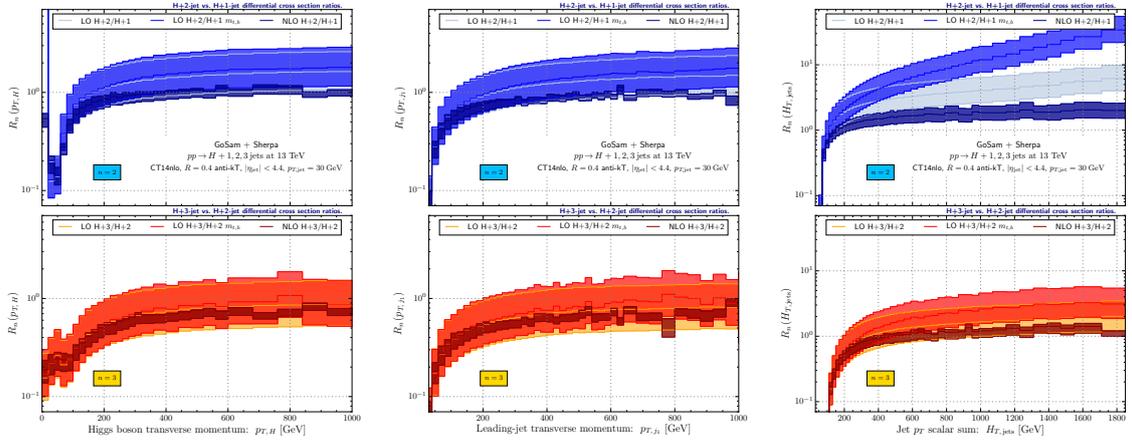

  \centering
  \begin{subfigure}{\linewidth}
    \includegraphics[width=0.327\textwidth]{./figs/%
      Multimeff_H+123_pT30eta44mult_higgs_pt_100vbf-E13.pdf}
    \hfill
    \includegraphics[width=0.327\textwidth]{./figs/%
      Multimeff_H+123_pT30eta44mult_jet_pT_1_200vbf-E13.pdf}
    \hfill
    \includegraphics[width=0.327\textwidth]{./figs/%
      Multimeff_H+123_pT30eta44mult_jets_ht_200-E13.pdf}
    \caption{\label{sfig:pts}%
      Comparison of effective theory predictions at LO and NLO with LO
      predictions (indicated by the extra `$m_{t,b}$' label) obtained in
      the full SM for various observables
      monitoring the transverse jet activity in the various Higgs boson
      plus jet production processes. The Higgs boson transverse momentum
      $p_{T,\,H}$, is shown on the left while the leading jet transverse
      momentum, $p_{T,\,j_1}$, and scalar $p_T$ sum of all jets,
      $H_{T,\,\mathrm{jets}}$, is presented in the middle pane and in
      the right pane, respectively. Note that the \Hj (green curves) and
      \Hjjj (red curves) predictions have been rescaled for better
      visibility. The smaller plots in the lower part of each panel show
      the ratios of the three different predictions normalized to the LO
      effective theory prediction. This is done separately for each of
      the \Hj, \Hjj and \Hjjj processes.}
  \end{subfigure}
  \vskip5mm
  \begin{subfigure}{\linewidth}
    \includegraphics[width=0.327\textwidth]{./figs/%
      Ratiomeff_H+123_pT30eta44mult_higgs_pt_100vbf-E13.pdf}
    \hfill
    \includegraphics[width=0.327\textwidth]{./figs/%
      Ratiomeff_H+123_pT30eta44mult_jet_pT_1_200vbf-E13.pdf}
    \hfill
    \includegraphics[width=0.327\textwidth]{./figs/%
      Ratiomeff_H+123_pT30eta44mult_jets_ht_200-E13.pdf}
    \caption{\label{sfig:ratio_n1_n_ptH_ptj1}%
      Differential \Hjj over \Hj ratios (upper row) and \Hjjj over
      \Hjj ratios (lower row) for the transverse momentum spectra of
      the Higgs boson, on the left, and the leading jet, in the
      middle, as well as the all-jets scalar $p_T$ sum, on the right.
      Each panel shows the ratios for the three theories considered
      here, the LO and NLO effective theory as well as the full theory
      at LO.}
  \end{subfigure}
  \caption{\label{fig:pts}%
    Various transverse momentum distributions and ratios between
    successive $n$-jet bins for the production of \Hnj
    ($n=1,2,3$) at the 13~\TeV LHC.}
  \vskip-7mm
\end{figure}

We start with the discussion of differential distributions relevant
for the LHC operated at a collider energy of 13~\TeV. In the figures
presented here, we compare the effective theory predictions at LO and
NLO with results obtained in the full SM. This provides us with a
direct comparison of the size of the different corrections. We can
decide more easily whether we need to pay attention to including the
NLO effects in the effective theory or the finite-mass effects based
on the full theory. We are also able to identify observables and/or
kinematical environments where it will be mandatory to incorporate
both effects in one way or another.

For the full theory calculations, we usually consider both of the
heaviest quarks running in the loop, i.e.~we take the top quark as
well as the bottom quark contributions into account. The associated
predictions will hence show the additional label `$m_{t,b}$' in the
figures. Depending on the specific observable, we will present all
three or two of the \Hnj predictions together in one plot (with $n=3$
being the maximum jet number). In the lower part of these plots, we
display for each jet bin separately, the ratios of the three different
types of predictions taken with respect to the corresponding LO result
in the effective theory. For example, in Figure~\ref{sfig:pts}, the
upper, middle and lower ratio plots show the different ratios based on
the predictions for \Hj, \Hjj and \Hjjj, respectively.

Transverse momentum distributions are known to receive significant
corrections that lead to a softening of the distribution for larger
$p_T$ values. We thus investigate this class of observables first, and
summarize most of our results in Figure~\ref{fig:pts}. For all three
processes, i.e.~for the production of \Hj, \Hjj and \Hjjj,
Figure~\ref{sfig:pts} displays next to each other the transverse
momentum distributions of the Higgs boson (left panel) and the hardest
jet (center panel), and the scalar sum of the transverse momenta of
all jets (right panel). These three distributions clearly show the
expected behaviour of $p_T$-tail softening. For small transverse
momenta (or small $H_T$), we find that the leading-order predictions
based on the effective theory are in very good agreement with the
respective leading-order predictions given by the SM. This is because
the heavy top-quark approximation works very well in this low-$p_T$
region. Furthermore, we see that this statement holds for all three
jet multiplicities considered here. Focusing on the pure $p_T$
distributions of Figure~\ref{sfig:pts}, we observe that the point at
which the effective theory approach starts to break down occurs around
Higgs boson or lead-jet values of $p_T=200~\GeV$ and is to a good
approximation independent of the jet multiplicity of the Higgs boson
production processes. This observation would support a rather simple
explanatory model, in which we assume that the resolution of the
effective vertex is mainly driven by a quantity, which is very
strongly correlated with the event's hardest single particle $p_T$.
The inner structure of the $ggH$ vertex will therefore be probed with
any interaction where the leading particle-$p_T$ exceeds the top-quark
mass. In \Hnj production, the hardest particle is either the Higgs
boson itself or the leading jet. This right away explains why the
breakdown occurs for both the $p_{T,\,H}$ and the $p_{T,\,j_1}$
spectra at the same scale. As we will argue further below, this
assumption for the main resolution driver seems to also work well for
the other examples of transverse observables discussed in this
section.

Above the breakdown scale, the deviation from the full SM predictions
becomes sizeable very rapidly, resulting in a strong suppression by a
factor of 10 at $p_T\sim1~\TeV$. Compared to this, the NLO corrections
in the effective theory lead to enhancements of the cross section,
which are distributed in a relatively uniform way.
Since we use $\hthatprime/2$ as our central scale, the
differential $K$-factor between the LO and NLO effective prediction
turns out to be flat for the production of \Hj~\cite{Greiner:2015jha},
whereas it has a non-trivial shape for \Hjj and \Hjjj production. The
latter two $K$-factors however approach $1$ for transverse momenta
that are about or larger than $600~\GeV$. It is therefore fair to say
that the NLO corrections turn into a subleading effect, already at
$p_T\sim400~\GeV$, as one has to contrast their behaviour with the
strong $p_T$ dependence of the finite-mass corrections.

We note that similar observations regarding finite-mass effects have
been made before. They have been pointed out in particular for the
$p_{T,\,H}$
distribution~\cite{Buschmann:2014sia,Frederix:2016cnl,Badger:2016bpw}. In
the one-jet case, we can use the leading-order exact statement that
$p_{T,\,H}=p_{T,\,j_1}$ ($=H_{T,\,\mathrm{jets}}$).  It is however
interesting to see that the differential ratios associated with
$p_{T,\,H}$ and $p_{T,\,j_1}$ (see the lower part of
Figure~\ref{sfig:pts}) are strikingly similar in their characteristics
even beyond the one-jet case. In addition, they are also very similar
among the different jet bins, suggesting that the relative $1/p^2_T$
scaling between the effective and full theory at LO can be applied in
a more universal manner (cf.~Section~\ref{sec:intro}). In fact if we
concentrate on the $p_{T,\,H}$ predictions, we observe that the
suggested scaling holds to a fairly good extent. For example, at
$p_{T,\,H}\approx400~\GeV$, the mass effects reduce the cross section
to roughly 60\% of the effective theory result. At $p_{T,\,H}>1~\TeV$,
this reduction then turns into an one-order of magnitude effect, which
fixes the related cross section ratio at a value of
\begin{equation}
  \dfrac{\tfrac{\mathrm{d}\sigma}{\mathrm{d}p_{T,\,H}}\big(p_{T,\,H}=1.0~\TeV\big)}
        {\tfrac{\mathrm{d}\sigma}{\mathrm{d}p_{T,\,H}}\big(p_{T,\,H}=0.4~\TeV\big)}
  \;\approx\;\frac{10\%}{60\%}\;=\;\frac{1}{6}~.
\end{equation}
The above number (as given by our computation) can be compared
with the number one expects from exploiting the relative scaling
property between the effective and full theory predictions. Based on
the additional suppression of the full result by two powers of
$p_{T,\,H}$, the expected value for the same cross section ratio
amounts to $(400~\GeV/1000~\GeV)^2=4/25$, which is very close to the
value extracted from the theory data. This result for the scaling does
not change much among the different jet bins because the three
ancillary plots in the lower part of Figure~\ref{sfig:pts} show that
the ratio between full and effective theory predictions only
marginally loses some of its steepness for an increasing final-state
multiplicity.

The cumulative characteristics of the $H_{T,\,\mathrm{jets}}$
observable shown in the right panel of Figure~\ref{sfig:pts} leads to
an amplification of the NLO effects in the effective theory (for
obvious reasons), while the full theory distributions in the multijet
cases fall off less severely at larger scales than they do for the
single-object $p_T$ spectra discussed above. We also notice that the
breakdown of the effective approach occurs at higher scales. In fact
an increasing jet multiplicity tames the finite-mass effects further,
i.e.~yields a weaker scaling and pushes the breakdown scale out to
larger values of $H_T$. The reason for these changes becomes clear by
looking at a fixed $H_{T,\,\mathrm{jets}}$ point, for example
$H_{T,\,\mathrm{jets}}=1~\TeV$. The transverse hardness is shared
among all jets, which also means that the leading jet appears
at a scale lower than $1~\TeV$. At this lower scale, the deviation of
the full theory $p_{T,\,j_1}$ prediction has not grown as large as for
exactly $1~\TeV$. This has to be reflected by the finite-mass
$H_{T,\,\mathrm{jets}}$ distribution, which therefore cannot fall as
quickly as the $p_{T,\,j_1}$ distribution.

As we are interested in the scaling properties of the finite-mass
effects, it is beneficial to study the ratios of successive
differential jet cross sections. Given a specific observable, we
define these ratios as
\begin{equation}\label{def:Rn}
  R_n(O)\;=\;
  \dfrac{\dfrac{\mathrm{d}\sigma}{\mathrm{d}O}
        \big(\Higgs\!+\!n\!\mbox{~jets}\big)}
        {\dfrac{\mathrm{d}\sigma}{\mathrm{d}O}
        \big(\Higgs\!+\!(n-1)\!\mbox{~jets}\big)}~.
\end{equation}
Figure~\ref{sfig:ratio_n1_n_ptH_ptj1} visualizes the $R_2$ and $R_3$
ratios (i.e.~the differential ratios for \Hjj/\Hj and \Hjjj/\Hjj cross
sections) for the transverse momentum observables discussed above. The
ratios are shown for each type of our predictions.
Figure~\ref{sfig:ratio_n1_n_ptH_ptj1} therefore supplements
Figure~\ref{sfig:pts} greatly, as it clearly exhibits the relative
importance of the respective subleading jet multiplicity at higher
scales, and the robustness of this feature under finite-mass effects.
For all three observables, we essentially find two regimes independent
of the type of the prediction: at low transverse scales, the $n$-jet
contribution is always significantly smaller than the ($n-1$)-jet
contribution, but rises quickly with increasing transverse scales.
This has already been pointed out in Ref.~\cite{Greiner:2015jha}.
Above a certain scale (which appears around $400~\GeV$ for $p_{T,\,H}$
and $p_{T,\,j_1}$, and around twice that scale for
$H_{T,\,\mathrm{jets}}$), one enters the saturation or scaling regime
where the $R_2$ and $R_3$ can be roughly described by a constant. This
indicates (and confirms our earlier statements) that nearly the same
scaling is in place in successive jet bins. The largest deviations
from this behaviour and between the different predictions can be found
for the $R_2(H_{T,\,\mathrm{jets}})$, which is no surprise again due
to the cumulative nature of the observable. The $R_2$ generally level
off at higher values than the $R_3$ where the inclusion of finite-mass
effects yields a slight increase of the respective LO effective
ratios. Again, the $R_2$ are somewhat more affected by this. The NLO
corrections to the effective predictions work in the opposite
direction. In all ratio distributions, they stabilize the constant
behaviour in the saturation regime.

\begin{figure}[t!]
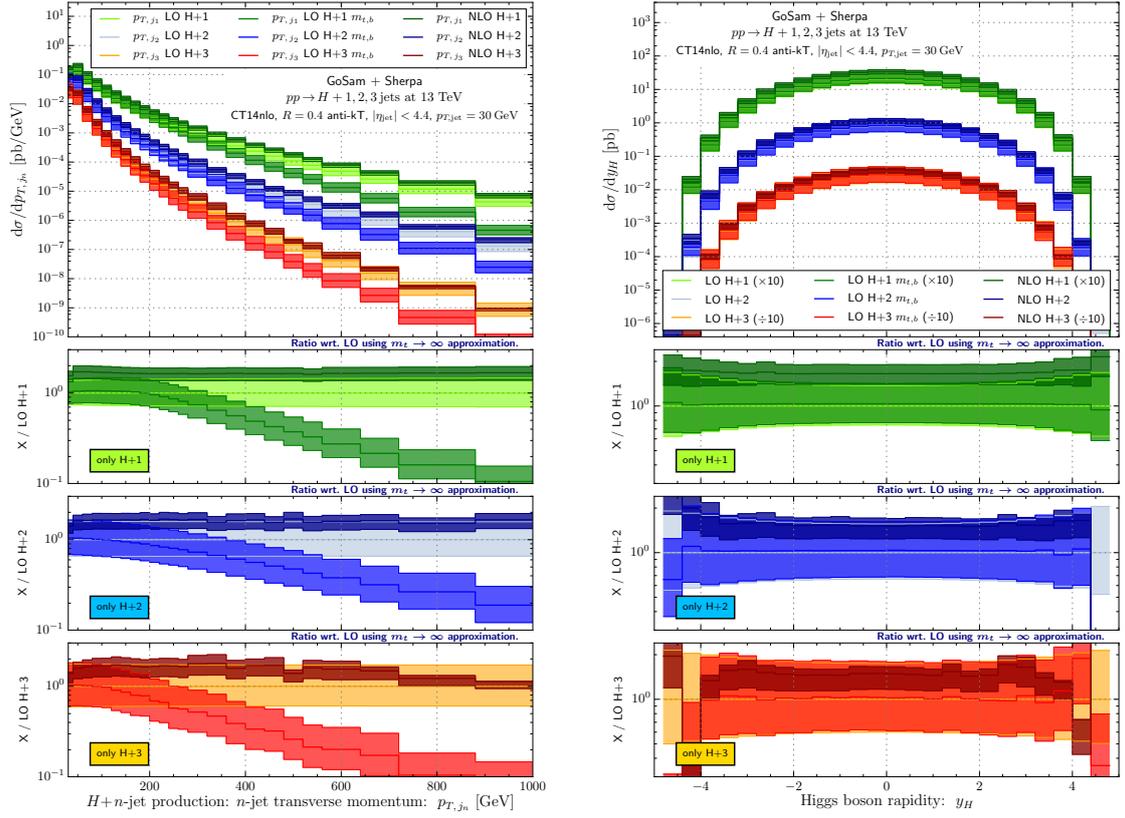

  \centering
  \includegraphics[width=0.49\textwidth]{./figs/%
    Multimeff_H+123_pT30eta44mult_wimpyjet_pT-E13.pdf}
  \hfill
  \includegraphics[width=0.49\textwidth]{./figs/%
    Multimeff_H+123_pT30eta44mult_higgs_y_50-E13.pdf}
  \caption{\label{fig:yh}%
    Comparison of the LO and NLO effective theory predictions with the
    LO predictions in the full SM for the transverse momentum of the
    $n$-th jet, $p_{T,\,j_n}$, in \Hnj production, on the left, and
    the Higgs boson rapidity, $y_H$, on the right. For the three modes
    of \Hj, \Hjj and \Hjjj production, each column of secondary plots
    shows the ratios of the three different predictions normalized to
    the one of the LO effective theory.}
\end{figure}

Figure~\ref{fig:yh} depicts, on the left hand side, the transverse
momentum distributions of the respective `wimpiest' jet in all three
\Hnj channels, i.e.~it shows the leading jet in \Hj, the second
leading jet in \Hjj and the third leading jet in \Hjjj production.
On the right hand side of the same figure, the rapidity distributions
of the Higgs boson are presented for the three cases of producing the
Higgs boson in association with one jet, two jets or three jets.
Compared to the observables discussed so far, the rapidity
distributions show a completely different behaviour. Here
the full theory and the effective theory agree throughout the entire
rapidity range. This is expected since the regions of the phase
space where the top-quark loop is resolved are more or less uniformly
distributed in rapidity, and their contribution is suppressed by at
least one order of magnitude (as shown by the $p_T$ spectra in
Figure~\ref{sfig:pts}) compared to the bulk of events, for which the
full and effective theory approaches agree. The NLO corrections
regarding the latter are sizeable although they mainly enhance the
cross section while leaving the shape more or less unaltered.

As can be seen in the left panel of Figure~\ref{fig:yh}, the effective
theory approach starts to deviate at even smaller values of the transverse
momentum, namely around $125~\GeV$ or $100~\GeV$, if one considers the
second leading jet in \Hjj production or the third leading jet in
\Hjjj production, respectively. This is a consequence of the $p_T$
ordering of the jets. In both cases there has to be a harder jet
present in the event that is distributed according to
$\mathrm{d}\sigma/\mathrm{d}p_{T,\,j_{n-1}}$. For the leading jet in
\Hjj and \Hjjj events, the deviation between the effective and full
theory results begins around $200~\GeV$ (see center panel in
Figure~\ref{sfig:pts}). The distributions for the second hardest jet
must therefore deviate around $(200-X)~\GeV$ where $X>0$, and
similarly, for \Hjjj final states, the third-jet distribution must
break down around $(200-X-Y)~\GeV$ where $Y>0$. Hence, the $p_T$
ordering of the jets translates into an ordering of breakdown scales.
In other words, if the superior jet does not resolve the heavy-quark
loop, the softer one will not do so at all.

\begin{figure}[t!]
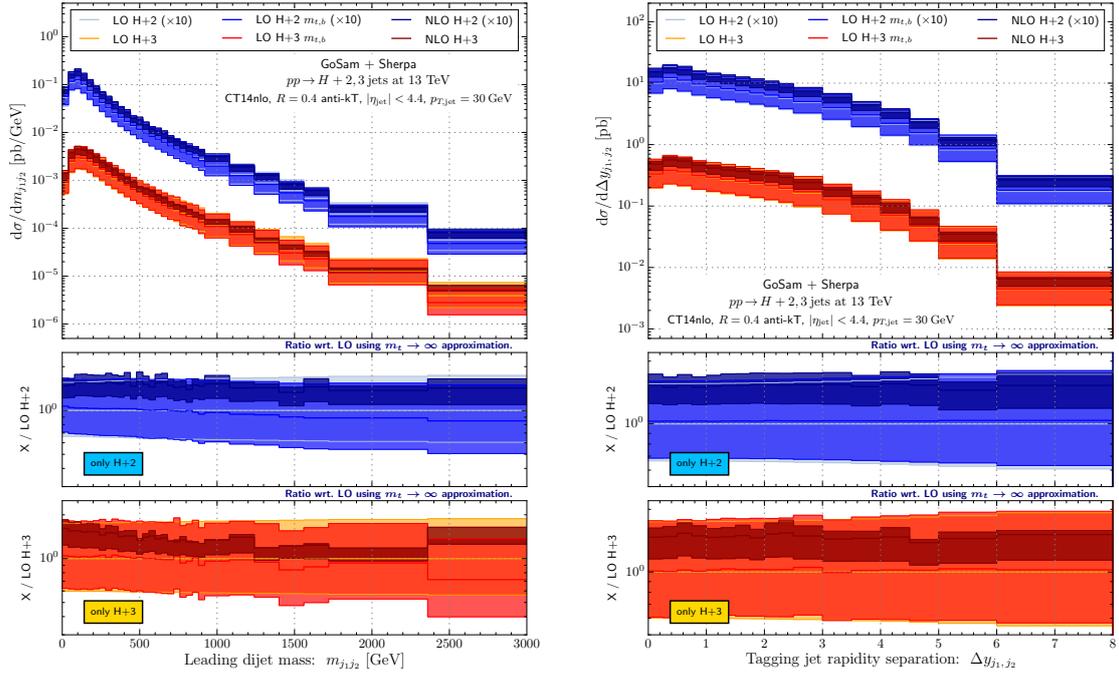

  \centering
  \includegraphics[width=0.49\textwidth]{./figs/%
    Multimeff_H+23_pT30eta44mult_jet_jet_mass_12_300-E13.pdf}
  \hfill
  \includegraphics[width=0.49\textwidth]{./figs/%
    Multimeff_H+23_pT30eta44mult_jet_jet_dy_12_40-E13.pdf}
  \caption{\label{fig:mjj}%
    Theoretical predictions based on the LO and NLO effective and the
    full approach at LO for the invariant mass (left) and the rapidity
    separation (right) of the two leading jets in \Hjj and \Hjjj
    production. The layout of the plots corresponds to that of
    Figure~\ref{sfig:pts}.}
\end{figure}

For the two-jet and three-jet processes, it is interesting to study
the invariant mass spectrum between the two hardest jets, which are
also the tagging jets in the $p_T$ jet-tagging scheme. This observable
plays an important role in the definition of kinematic constraints for
VBF analyses. The other key observable needed in VBF studies is the
rapidity separation of the same pair of jets. Both distributions are
shown in Figure~\ref{fig:mjj} for \Hjj and \Hjjj final states. For the
invariant mass distribution, one observes only a mild deviation
between the full and the effective theory predictions, which becomes
more pronounced towards the higher end of the kinematical range. As a
matter of fact, large invariant tag-jet masses are not only generated
in events with hard tagging jets. The geometry of the momentum flow is
an important criteria for the production of dijet masses. For example,
in a situation where the two jets appear in a back-to-back
configuration, large invariant masses can emerge despite the absence
of energetic jets. In these cases, the effect of the heavy quarks is
significantly reduced, and the effective theory therefore can be used
to describe this observable accurately.

For the rapidity separation between the two tagging jets, shown on the
right hand side in Figure~\ref{fig:mjj}, the situation is similar to
that of the rapidity distribution of the Higgs boson discussed above.
It is therefore clear that there is a good agreement between the
effective theory and the full theory predictions for this observable.

\begin{figure}[t!]
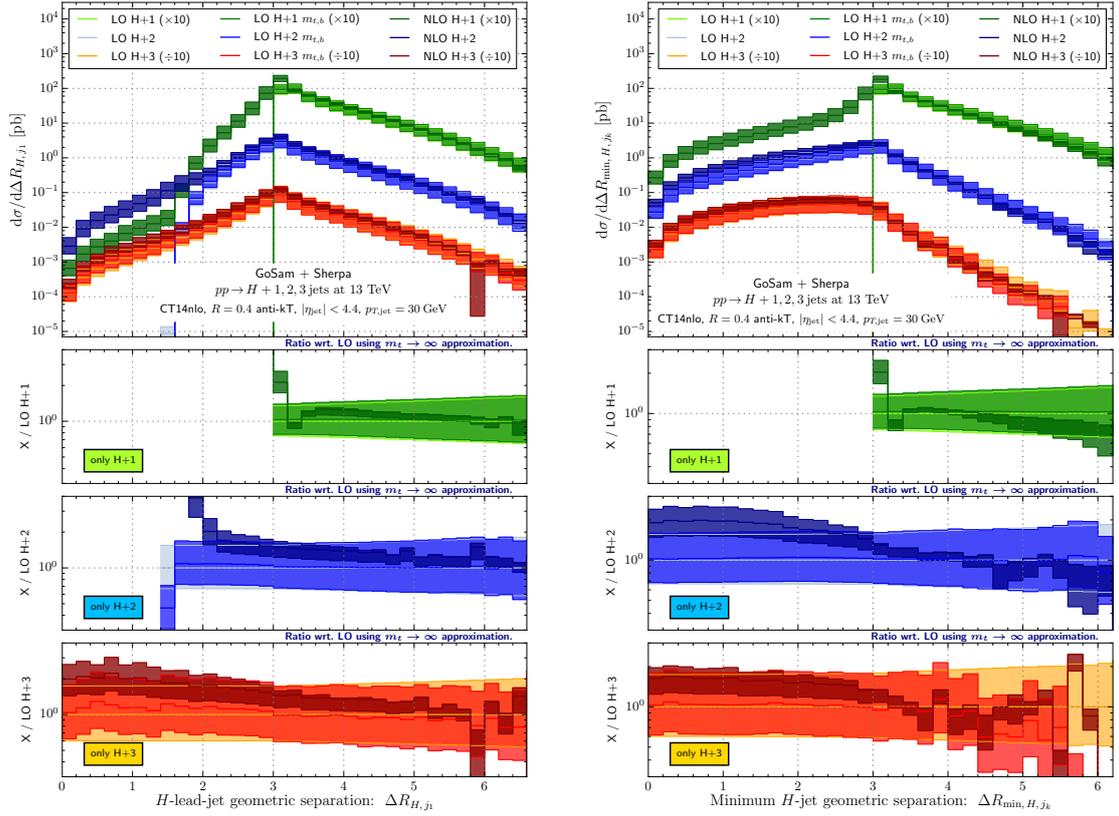

  \centering
  \includegraphics[width=0.49\textwidth]{./figs/%
    Multimeff_H+123_pT30eta44mult_higgs_jet_dR_1_35-E13.pdf}
  \hfill
  \includegraphics[width=0.49\textwidth]{./figs/%
    Multimeff_H+123_pT30eta44mult_higgs_jet_dR_min-E13.pdf}
  \caption{\label{fig:dRhj}%
    Theoretical predictions based on the LO and NLO effective and the
    full approach at LO for the geometric separation, in terms of
    radial distances, between the Higgs boson and the leading
    transverse momentum jet (left panel), and between the Higgs boson
    and the closest jet in the event (right panel). The layout of the
    plots corresponds to one used in Figure~\ref{sfig:pts}.}
\end{figure}

Figure~\ref{fig:dRhj} shows on the left hand side the radial
separation $\Delta R_{H,\,j_1}$ between the Higgs boson and the
leading jet. On the right hand side, we display the smallest of the
radial distances between the Higgs boson and any of the jets in the
event, $\Delta R_{\mathrm{min},\,H,\,j_k}$. In \Hj production at LO,
the Higgs boson and the only jet present in the event are forced into
a back-to-back configuration. The $\Delta R_{H,\,j_1}$ distribution
has therefore a natural cut-off at $\pi$, where it also peaks. At NLO,
the presence of a second jet, which can become unresolved, opens up
the previously kinematically forbidden range between $0$ and $\pi$.
The \Hjj distribution at LO also has a kinematical constraint owing to
the presence of two jets that must be resolved. The radial distance
between the Higgs boson and the leading jet can therefore not be
smaller than the minimal azimuthal angle of $\Delta\phi=\pi/2$. The
presence of at least a third jet (in \Hjj final states at NLO and
\Hjjj final states) allows one to finally populate the full
kinematical spectrum. From the distributions, it is however clear that
the Higgs boson preferably recoils against the leading jet, because
independent of the jet multiplicity, the distributions are all peaked
at $\Delta R_{H,\,j_1}=\pi$. The finite-mass effects are only very
mild in the \Hj and \Hjj case. In the \Hjjj case, they give a small
correction that slightly increases the cross section at small radial
separation, and decreases it at values larger than $\Delta
R_{H,\,j_1}=\pi$.  This is a consequence of $\Delta R$ being derived
from the rather robust variables $\Delta y$ and $\Delta\phi$.

\begin{figure}[t!]
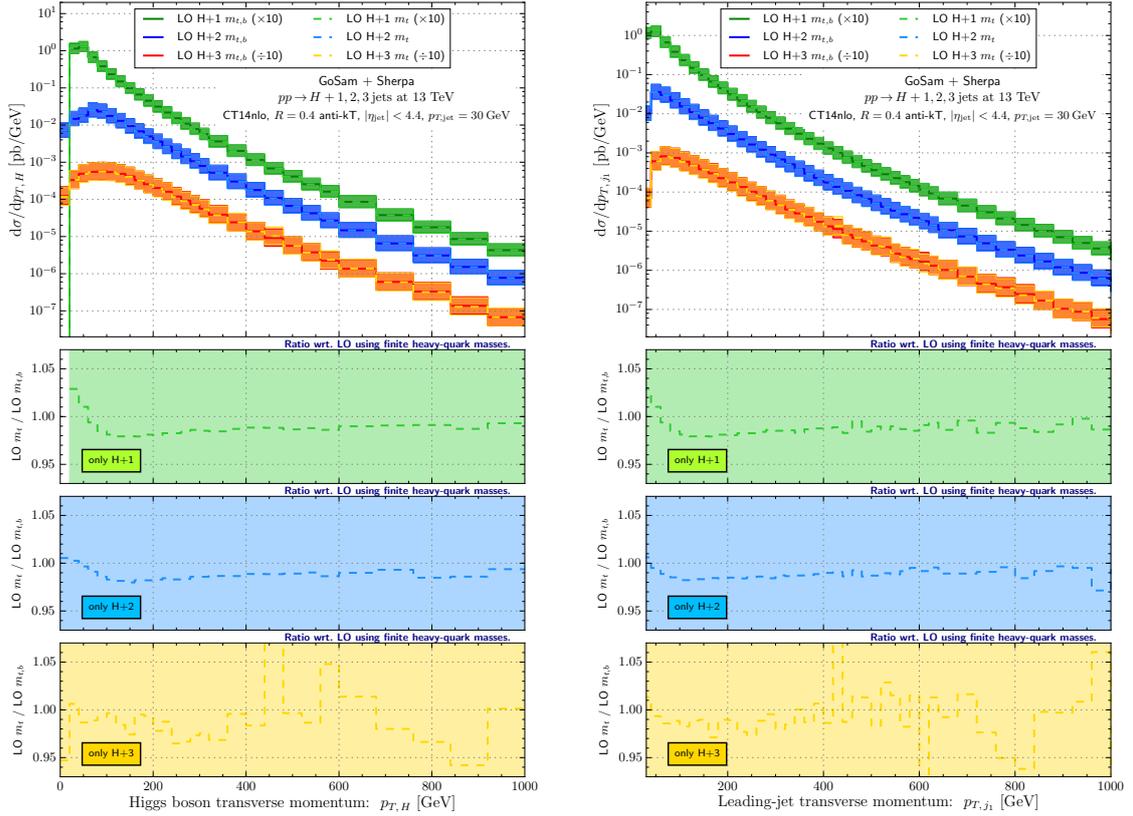

  \centering
  \includegraphics[width=0.49\textwidth]{./figs/%
    Mtbmeff_H+123_pT30eta44mult_higgs_pt_100vbf-E13.pdf}
  \hfill
  \includegraphics[width=0.49\textwidth]{./figs/%
    Mtbmeff_H+123_pT30eta44mult_jet_pT_1_200vbf-E13.pdf}
  \caption{\label{fig:mtb_mt_pts}%
    Transverse momentum distributions of the Higgs boson (left) and
    leading transverse momentum jet (right). The curve show
    predictions in the full standard model with and without additional
    bottom-quark loops for the three studied jet multiplicity. The
    width of the band represents the scale uncertainty. The lower
    plots show the ratio between the two central values. Because of
    the very small differences the edges of the scale uncertainty band
    is not visible in the lower plots.}
\end{figure}

For the same reason mentioned previously, the minimal radial
separation between the Higgs boson and a jet has a kinematic edge at
$\pi$ in \Hj production at LO. In all other cases, the distribution is
spread over the entire kinematical range. It is interesting to notice
that, contrary to the plot on the left, in the \Hjj and \Hjjj final
states, the distributions flatten out for $\Delta R_{\mathrm{min},\,H,\,j_k}$
values between $2$ and $\pi$. Based on these and earlier findings, a
typical event where the Higgs boson is produced in association with
several jets will likely have these features: the Higgs boson will
tend to recoil against the leading jet, but clearly, as the
multiplicity increases, it will occur more often in company of a
close, rather soft jet. Finally we remark on the good agreement
between the predictions from the effective theory and the predictions
including the mass corrections. Only for \Hjjj production, and for
large radial separations between the jets do the two curves start to
deviate. In contrast to this, the NLO corrections in the effective
approach have a significantly larger impact, and this statement
extents to all 2-body correlations considered in this section.

\subsection{The case of massless bottom quarks}\label{sec:bottom_vs_top}

In this section we compare the predictions for differential
distributions in the full theory with and without the $b$-quark loop
contribution, with the aim to assess the effects of neglecting the
bottom quark contribution. In Table~\ref{tab:xsec_ggf} and
Figure~\ref{fig:inclxs} above we already commented the impact on the
total cross sections, and observed a change in the sign of these
contribution between the \Hj predictions and \Hjj and \Hjjj
predictions. This can now be quantified better at the differential level.

\begin{figure}[t!]
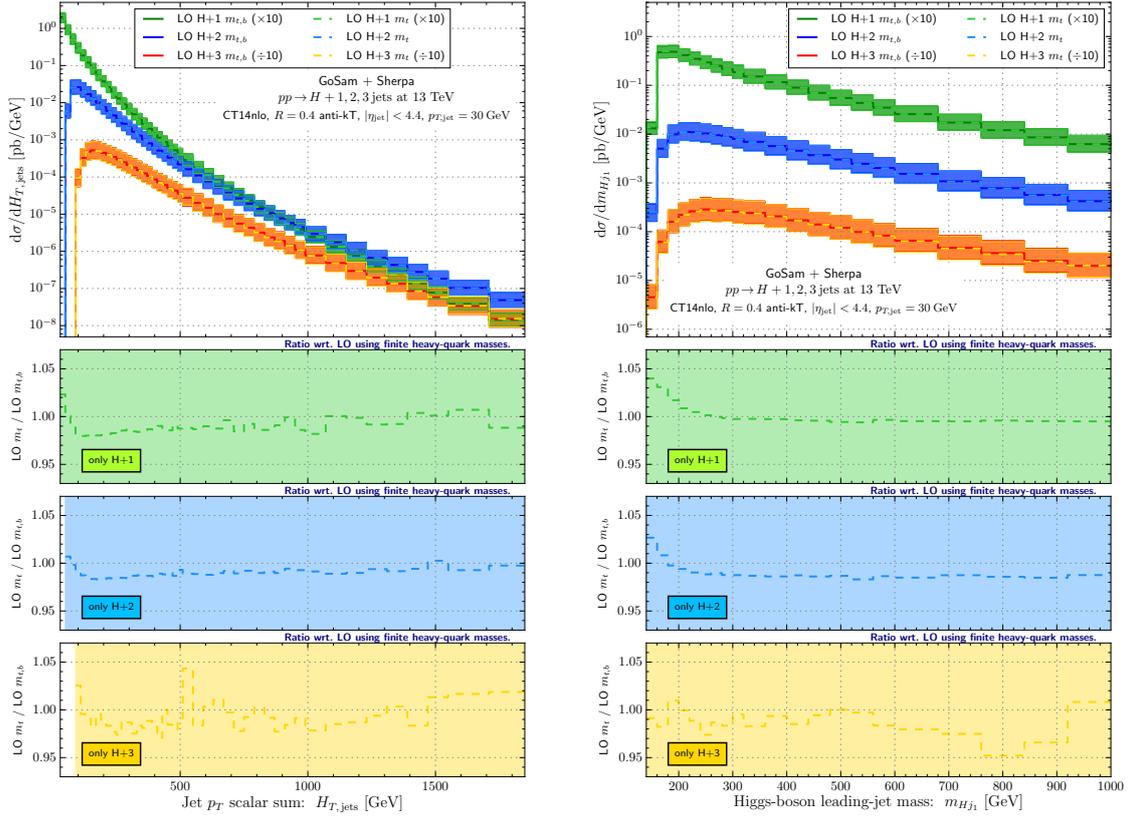

  \centering
  \includegraphics[width=0.49\textwidth]{./figs/%
    Mtbmeff_H+123_pT30eta44mult_jets_ht_200-E13.pdf}
  \hfill
  \includegraphics[width=0.49\textwidth]{./figs/%
    Mtbmeff_H+123_pT30eta44mult_higgs_jet_mass_1_100-E13.pdf}
  \caption{\label{fig:mtb_mt_hts}%
    Same as Fig.~\ref{fig:mtb_mt_pts}, but for the scalar sum of the jet
    transverse momenta and the invariant mass of the Higgs boson and
    the leading transverse momentum jet.}
\end{figure}

\begin{figure}[t!]
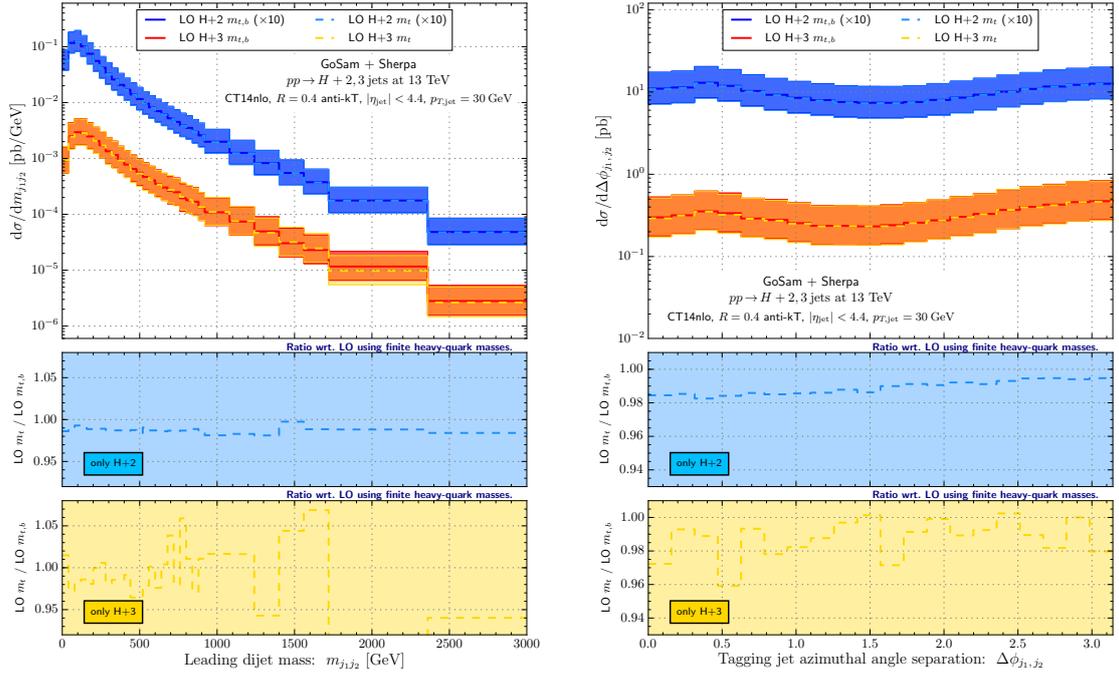

  \centering
  \includegraphics[width=0.49\textwidth]{./figs/%
    Mtbmeff_H+23_pT30eta44mult_jet_jet_mass_12_300-E13.pdf}
  \hfill
  \includegraphics[width=0.49\textwidth]{./figs/%
    Mtbmeff_H+23_pT30eta44mult_jet_jet_dphi_12_20-E13.pdf}
  \caption{\label{fig:mtb_mt_jj}%
    Same as Fig.~\ref{fig:mtb_mt_pts}, but for the invariant mass of the
    leading dijet system (left) and their azimuthal angle separation
    (right).}
\end{figure}

For this we focus our attention on a few selected observables. In
Figure~\ref{fig:mtb_mt_pts} we show the Higgs boson and the leading
jet transverse momenta, in Figure~\ref{fig:mtb_mt_hts} we compare the
scalar sum of the jet transverse momenta $H_{T,\mathrm{jets}}$ and the
invariant mass of Higgs boson and leading transverse momentum jet,
whereas in Figure~\ref{fig:mtb_mt_jj} we plot the invariant mass of
the tagging jets and their azimuthal angle difference
$\Delta\phi_{j1,j2}$. All the plots show the corresponding observable
as well as ratio plots for the three different jet-multiplicities. The
ratio is given by the result in which only top-quark loops are
considered, divided by the predictions where both top- and
bottom-quark contributions are taken into account. The color shaded
areas denote the scale uncertainty. For a better visibility we do not
show the full scale uncertainty band in the ratios, but rather zoom in
around the central scale.

It is clearly visible that the scale uncertainty outweighs the bottom
mass effects by far, for all the considered observables. The size of
the effects strongly depends on the observable but never exceeds five
percent. In general, the bottom-quark mass effects are most visible in
the observables involving transverse momenta and sums thereof and in
invariant masses involving the Higgs. It is however interesting to
observe which of the two predictions is larger as a function of the
kinematical region considered. The largest effects can be observed in
the soft region of the observables. This is to be expected, since
especially when the kinematical scales involved are not too large
compared to the bottom-quark mass, bottom-quark loops can lead to
sizable corrections to the predictions in which only the top quark is
considered. Far away from these kinematic regions the bottom quark can
be considered massless. Furthermore, as already discussed in section
\ref{sec:total_xsec}, the size and sign of the effect depends on the
jet multiplicity. This can be seen for instance in
Figure~\ref{fig:mtb_mt_hts}. The destructive interference for \Hj at
the level of the total cross section stems from the soft region,
whereas the net contribution becomes positive in regions where the
b-quark can be considered massless. For \Hjj and \Hjjj the destructive
interference is considerably reduced, leading to an increase of the
total cross section when bottom-quark loops are taken into
account. For angular variables these contributions are instead flat,
over the whole kinematical range. This is expected since the effects
are uniformly distributed in these variables as already discussed in
the previous section.

\subsection{VBF measurements at the LHC}

The production of a Higgs boson in association with two or more jets
in the gluon-fusion channel is also the main background to the VBF
production channel. Since the latter has a very characteristic
topological signature, in which two jets are produced mainly at high
rapidities with a large invariant mass and a large azimuthal separation,
leaving little jet activity in the central region of the detector,
this channel can be enhanced with respect to the background by
additional cuts, similar to the one of Eq.~\ref{cuts:vbf}. In this
section we investigate the impact of the mass effects on the
gluon-fusion predictions when these additional cuts are applied.

\begin{table}[t!]
\centering\small
\begin{tabular}{l|c|c}\hline
  \hline
  Numbers in [pb]
  & $\mathrm{H}\!+\!2$ jets\phantom{\Big|}
  & $\mathrm{H}\!+\!3$ jets\phantom{\Big|} \\
  \hline
  $\sigma_{\mathrm{LO,\,eff.}}$  \phantom{\Big|} & $0.397^{+64\%}_{-36\%}$ & $0.166^{+82\%}_{-42\%}$ \\
  $\sigma_{\mathrm{NLO,\,eff.}}$ \phantom{\Big|} & $0.584^{+10\%}_{-19\%}$ & $0.231^{+5\%}_{-22\%}$  \\
  $\sigma_{\mathrm{LO},\,m_{t,b}}$  \phantom{\Big|} & $0.404^{+65\%}_{-37\%}$ & $0.167^{+82\%}_{-42\%}$ \\
  $\sigma_{\mathrm{LO},\,m_{t}}$ \phantom{\Big|} & $0.398^{+65\%}_{-37\%}$ & $0.165^{+82\%}_{-42\%}$ \\
  \hline
  \hline
  \end{tabular}
  \caption{\label{tab:xsec13vbf}%
    Total inclusive cross sections for the production of a Higgs boson
    in association with two or three jets at LO and NLO in QCD in the
    effective theory and at LO in the full SM for massive top- and
    bottom-quarks and for massive top quarks only. Numbers are
    reported for center-of-mass energies of 13~\TeV. In addition to
    the jet transverse momentum cut of $p_{T,\,\mathrm{jet}}>30~\GeV$,
    we demand $m_{j_1 j_2}\,>\,400~\GeV$ and $\left|\Delta
    y_{j_1,\,j_2}\right|\,>\,2.8$. The uncertainty estimates are
    obtained from standard scale variations.}
\end{table}

As already demonstrated by Figure~\ref{fig:mjj}, both the $m_{j_1j_2}$
variable and the $\Delta y_{j_1,j_2}$ variable are almost unaffected
by mass corrections. We can therefore expect that at the level of the
total cross section the pattern stays similar as the case without VBF
cuts. The total cross sections reported in Table~\ref{tab:xsec13vbf}
confirm this. Also the effect of the massive bottom-quark loops is
very small leading to changes in range of $1-2\%$. Therefore, for the
remainder of this section, we will not discriminate between top-quark
only and combined top-quark and bottom-quark effects, instead always
include both massive quark loop contributions at the same time.

\begin{figure}[t!]
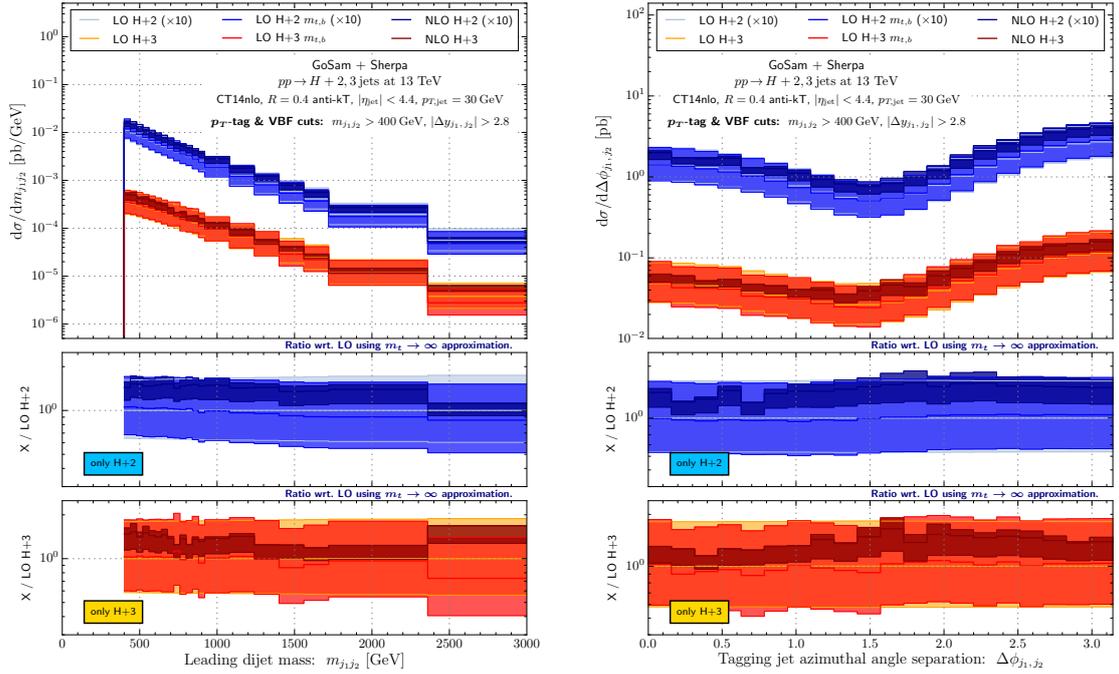

  \centering
  \includegraphics[width=0.49\textwidth]{./figs/%
    Multimeff_H+23_pT30eta44mult-vbf_jet_jet_mass_12_300-E13.pdf}
  \hfill
  \includegraphics[width=0.49\textwidth]{./figs/%
    Multimeff_H+23_pT30eta44mult-vbf_jet_jet_dphi_12_20-E13.pdf}
  \caption{\label{fig:vbf_mjj_dphijj}%
    Leading dijet invariant mass distribution (left) and azimuthal
    angle separation of the tagging jets (right) passing the VBF
    selection cuts.}
\end{figure}

Two of the key observables in the VBF scenario are the invariant mass
distribution of the tagging jets $m_{j_1j_2}$ and their azimuthal
angular separation $\Delta\phi_{j_1,j_2}$ . These two observables are
shown in Figure~\ref{fig:vbf_mjj_dphijj}. The comparison between the
full theory and the effective theory predictions reveals a good
agreement between the two, indicating that mass effects are rather small.

\begin{figure}[t!]
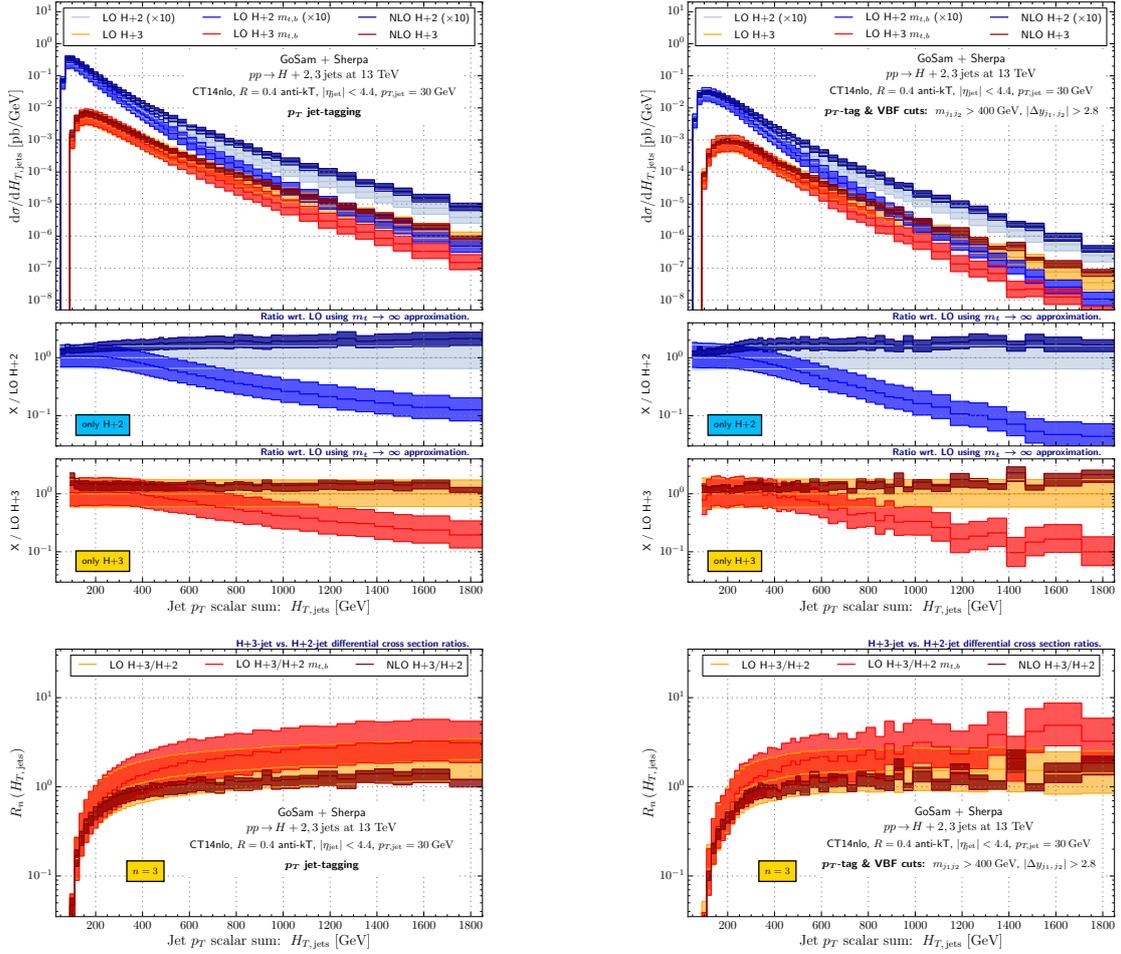

  \centering
  \includegraphics[width=0.45\textwidth]{./figs/%
    Multimeff_H+23_pT30eta44mult_jets_ht_200-E13.pdf}
  \hfill
  \includegraphics[width=0.45\textwidth]{./figs/%
    Multimeff_H+23_pT30eta44mult-vbf_jets_ht_200-E13.pdf}
  \\[-2mm]
  \includegraphics[width=0.45\textwidth]{./figs/%
    Ratiomeff_H+23_pT30eta44mult_jets_ht_200-E13.pdf}
  \hfill
  \includegraphics[width=0.45\textwidth]{./figs/%
    Ratiomeff_H+23_pT30eta44mult-vbf_jets_ht_200-E13.pdf}
  \caption{\label{fig:vbfHTjets}%
    The distribution of the scalar transverse momentum sum of all jets
    in H+2-jet and H+3-jet events before and after passing the VBF
    selection cuts.}
\end{figure}

These effects become, however, considerably larger when one considers
observables that have already been seen to be sensitive to heavy-quark
loops in the previous sections. In Figure~\ref{fig:vbfHTjets} we show
the scalar sum of the transverse momenta $H_{T,\mathrm{jets}}$ of the
jets before (left hand side)  and after applying the VBF cuts (right
hand side).  The effect is clearly visible when comparing the ratio
plots normalized to the LO in the effective theory: after VBF cuts the
mass effects become more severe, leading to larger deviations of the
effective theory from the full theory for high transverse momenta. In
the lowest row we show the differential ratio between \Hjjj and \Hjj,
which remains roughly unchanged.

\begin{figure}[t!]
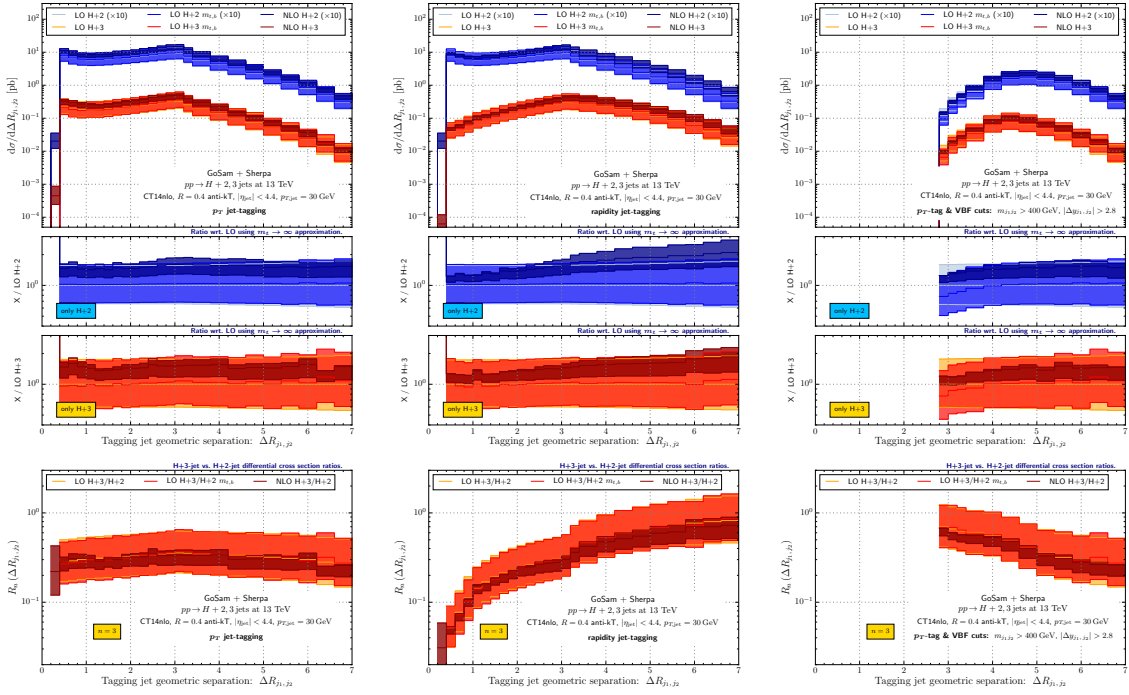

  \centering
  \includegraphics[width=0.327\textwidth]{./figs/%
    Multimeff_H+23_pT30eta44mult_jet_jet_dR_12_35-E13.pdf}
  \hfill
  \includegraphics[width=0.327\textwidth]{./figs/%
    Multimeff_H+23_pT30eta44mult-tag_jet_jet_dR_12_35-E13.pdf}
  \hfill
  \includegraphics[width=0.327\textwidth]{./figs/%
    Multimeff_H+23_pT30eta44mult-vbf_jet_jet_dR_12_35-E13.pdf}
  \\[-2mm]
  \includegraphics[width=0.327\textwidth]{./figs/%
    Ratiomeff_H+23_pT30eta44mult_jet_jet_dR_12_35-E13.pdf}
  \hfill
  \includegraphics[width=0.327\textwidth]{./figs/%
    Ratiomeff_H+23_pT30eta44mult-tag_jet_jet_dR_12_35-E13.pdf}
  \hfill
  \includegraphics[width=0.327\textwidth]{./figs/%
    Ratiomeff_H+23_pT30eta44mult-vbf_jet_jet_dR_12_35-E13.pdf}
  \caption{\label{fig:vbfRjj}%
    Geometric separation of the tagging jets in \Hjj and \Hjjj events
    when applying only baseline cuts in the $p_T$-jet tagging (left
    panel), for baseline cuts with rapidity ordering (middle panel)
    and when applying additional VBF cuts in the $p_T$ jet-tagging
    scheme (right panel).}
\end{figure}

Another important observable is the radial separation between the two
tagging jets $\Delta R_{j_1,j_2}$. This observable is considered in
Figure~\ref{fig:vbfRjj}. On the left and in the middle column we show
this observable before applying VBF cuts. The first plot shows the
observable for a $p_T$ jet-tagging, whereas in the second plot we
adopt a $y$ jet-tagging strategy. Although NLO effects in the
effective theory lead to substantial differences between the two
tagging schemes, the two leading order results in the full and the
effective theory agree very well for both tagging schemes and for both
multiplicities. At least at leading order the choice of the tagging
scheme is therefore insensitive to mass effects. The plot on the
right hand side shows the observable using the original $p_T$
jet-tagging but after applying VBF cuts. In this case mass corrections
have an impact especially for $\Delta R\approx3$ , where deviations
can become larger than $20\%$. When the two jets are not too far in
radial distance, the Higgs boson must in general be harder and recoil
against them. This explains the discrepancy between the two LO
predictions. The bottom plots show the ratio between the two jet
multiplicities. As expected, both the tagging scheme as well as the
VBF cuts have a significant impact on this ratio. However, heavy-quark
mass effects do not lead to deviations from the prediction of the
effective theory.

\subsection{FCC predictions for 100 TeV collisions}

In view of a possible future circular collider operating at a
center-of-mass energy of $100$~\TeV, we also investigate the mass
effects for such a high collider energy. Values for the total cross
section at the different multiplicities were already presented and
discussed above in Section~\ref{sec:total_xsec} and
Table~\ref{tab:xsec_ggf}. In this section we focus on differential
distributions.

\begin{figure}[t!]
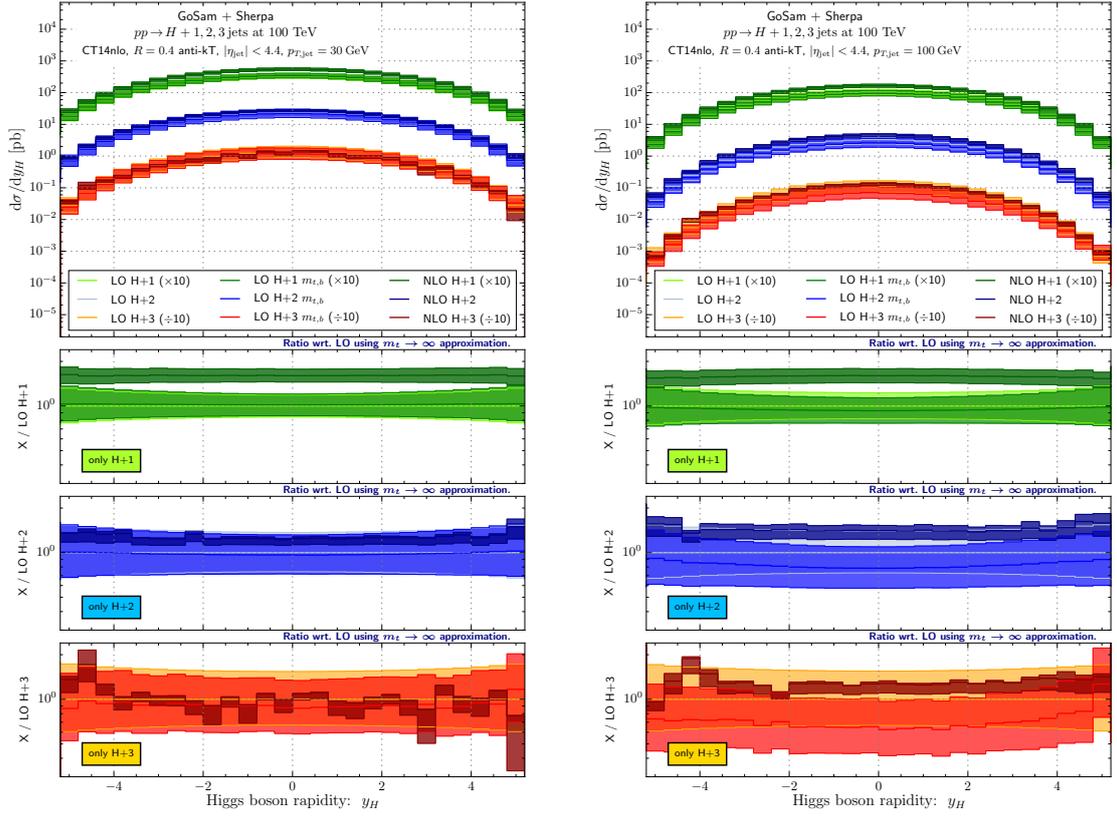

  \centering
  \includegraphics[width=0.49\textwidth]{./figs/%
    Multimeff_H+123_pT30eta44mult_higgs_y_50-E100.pdf}
  \hfill
  \includegraphics[width=0.49\textwidth]{./figs/%
    Multimeff_H+123_pT100eta44mult_higgs_y_50-E100.pdf}
  \caption{\label{fig:fccraps}%
    The Higgs boson rapidity distribution for a low and high jet $p_T$
    threshold.}
\end{figure}

In Figure~\ref{fig:fccraps} we investigate the impact of the mass
effects depending on the cut on the jets transverse momentum by
looking at the Higgs boson rapidity distribution. On the left the
transverse momentum cut on the jets is $p_{T,\,\mathrm{jet}}>30$~\GeV,
whereas on the right it is increased to
$p_{T,\,\mathrm{jet}}>100$~\GeV. At $13$~\TeV the leading order
contributions of the full and the effective theory agree very well and
we find the same good agreement at 100~\TeV for the looser transverse
momentum cut. Requiring a minimum transverse momentum of $100$~\GeV
leads to visible deviations between full and effective theory. This is
clearly related to the fact that the bulk of the cross section comes
from the softest allowed region of the phase space, where the mass
effects play only a very minor role. Increasing the $p_T$ threshold,
however, cuts away this large and mass-insensitive part of the cross
section and the remaining contribution is much more affected by mass
effects.

\begin{figure}[t!]
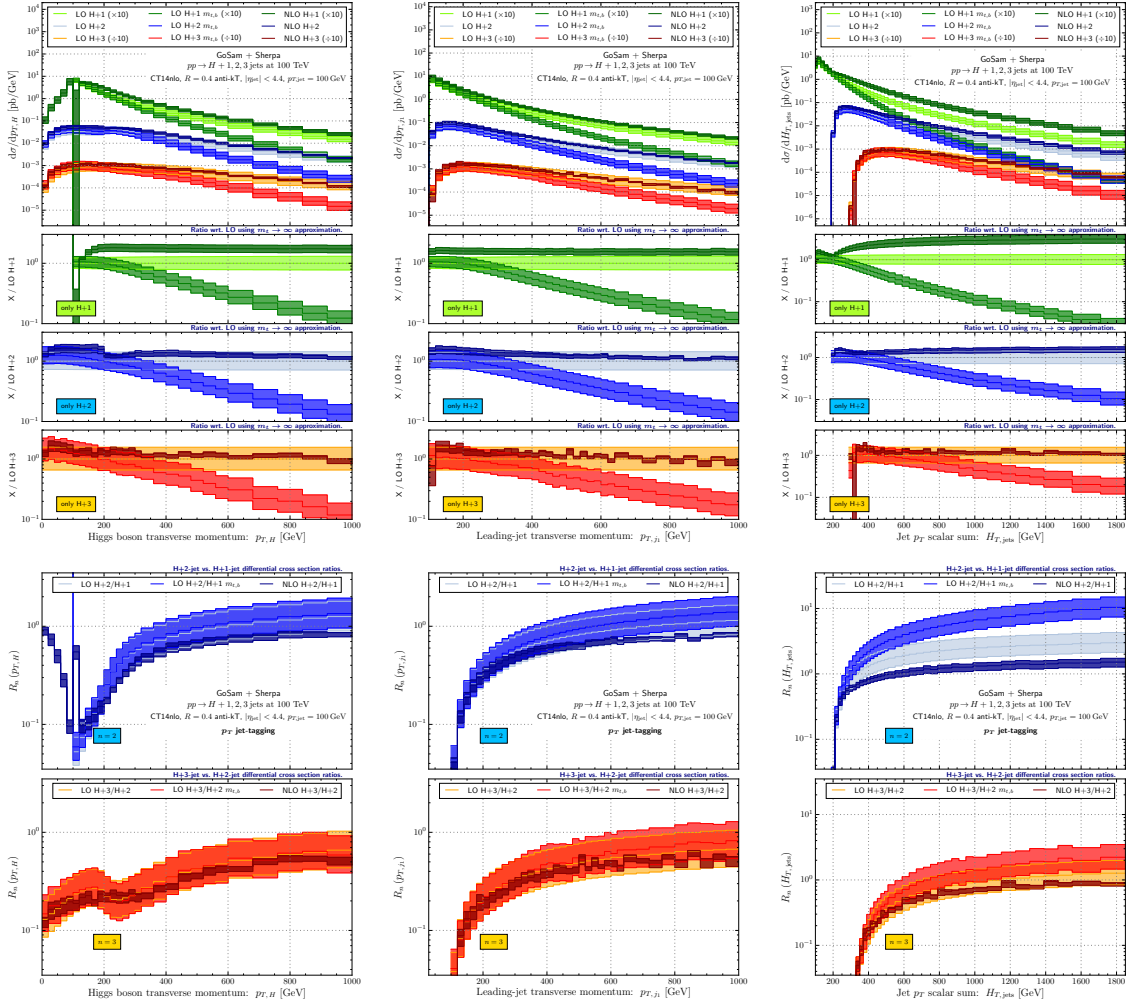

  \centering
  \includegraphics[width=0.327\textwidth]{./figs/%
    Multimeff_H+123_pT100eta44mult_higgs_pt_100vbf-E100.pdf}
  \hfill
  \includegraphics[width=0.327\textwidth]{./figs/%
    Multimeff_H+123_pT100eta44mult_jet_pT_1_200vbf-E100.pdf}
  \hfill
  \includegraphics[width=0.327\textwidth]{./figs/%
    Multimeff_H+123_pT100eta44mult_jets_ht_200-E100.pdf}
  \\[-2mm]
  \includegraphics[width=0.327\textwidth]{./figs/%
    Ratiomeff_H+123_pT100eta44mult_higgs_pt_100vbf-E100.pdf}
  \hfill
  \includegraphics[width=0.327\textwidth]{./figs/%
    Ratiomeff_H+123_pT100eta44mult_jet_pT_1_200vbf-E100.pdf}
  \hfill
  \includegraphics[width=0.327\textwidth]{./figs/%
    Ratiomeff_H+123_pT100eta44mult_jets_ht_200-E100.pdf}
  \caption{\label{fig:fccpts}%
    Various transverse momentum distributions for the three Higgs
    boson plus $n$-jet bins and their associated successive jet bin
    ratios.}
\end{figure}

In Figure~\ref{fig:fccpts} we show the transverse momentum of the
Higgs and the leading jet as well as $H_T$ of the jets with a $p_T$
cut on the jets of $100$~\GeV. Owing to the increase in the cross
section and the possibility to produce much harder jets and Higgs
bosons, all these observables suffer from large mass effects, which
for transverse momenta larger than $1$~\TeV lead to corrections which
are bigger than one order of magnitude. The lowest panel shows the inclusive
differential \Hjj/\Hj and \Hjjj/\Hjj ratios for the three
different observables. These ratios remain unchanged for the
transverse momentum of the Higgs boson, meaning that the relative
importance of higher multiplicity contributions is stable under mass
effect corrections, and we also see an only very mild deviation for the
transverse momentum of the leading jet.
However, for $H_{T}$ the ratio increases when passing from the
effective theory predictions to the full SM. The massive quark loop
effects are therefore stronger in the high transverse momentum tails
for the lower multiplicities than for the higher ones.

\begin{figure}[t!]
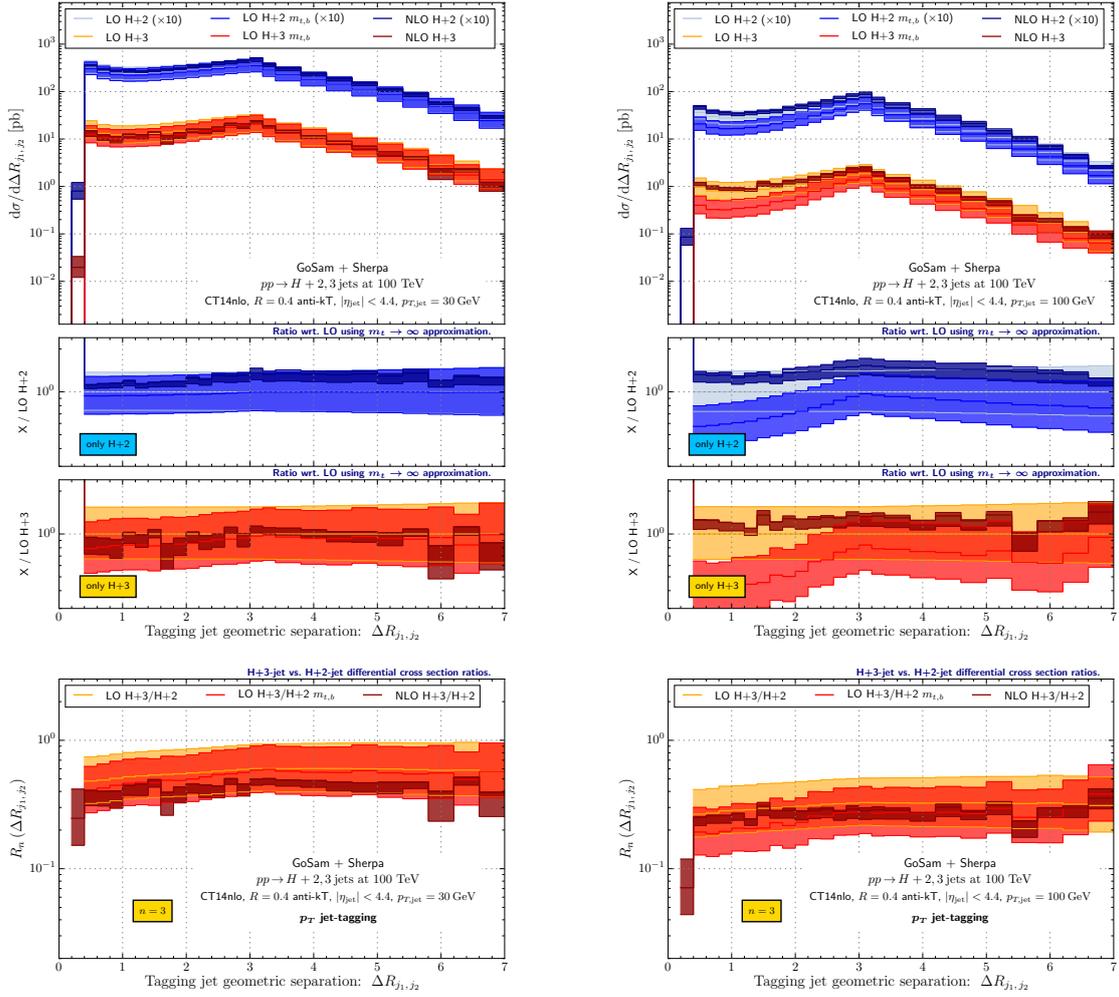

  \centering
  \includegraphics[width=0.47\textwidth]{./figs/%
    Multimeff_H+23_pT30eta44mult_jet_jet_dR_12_35-E100.pdf}
  \hfill
  \includegraphics[width=0.47\textwidth]{./figs/%
    Multimeff_H+23_pT100eta44mult_jet_jet_dR_12_35-E100.pdf}
  \\[-2mm]
  \includegraphics[width=0.47\textwidth]{./figs/%
    Ratiomeff_H+23_pT30eta44mult_jet_jet_dR_12_35-E100.pdf}
  \hfill
  \includegraphics[width=0.47\textwidth]{./figs/%
    Ratiomeff_H+23_pT100eta44mult_jet_jet_dR_12_35-E100.pdf}
  \caption{\label{fig:fccRjj}%
    Geometric separation between the leading jets for a low and high
    jet threshold and the associated $R_3$ ratios using $p_T$
    jet-tagging.}
\end{figure}

Figure~\ref{fig:fccRjj} shows again the radial separation between the
two leading jets $\Delta R_{j_1,j_2}$, this time for 100~\TeV center
of mass energy. The two plots show the impact of the finite mass
corrections when the minimum transverse momentum threshold is raised
from $p_{T,\,\mathrm{jet}}>30$~\GeV to
$p_{T,\,\mathrm{jet}}>100$~\GeV. The very small differences between
the effective theory predictions and the full SM curve at 13~\TeV
(Figure~\ref{fig:vbfRjj} -- left) become larger, especially for \Hjjj,
when increasing the collider energy, even if the cuts are kept
equal. On the lowest ratio of the left plot we observe a non-trivial
shape of the mass corrections, which are minimal when the separation
is about $\Delta R_{j_1,j_2}\approx\pi$. For smaller separations the
two leading jets must be close in the $(y,\phi)$-plane and combine
such that the momentum flow through the effective vertex is increased, leading
to a break down of the effective theory prediction. The tiny
variations in the shape of the mass corrections dramatically increase
when the transverse momentum of the jets is required to be above
$100$~\GeV. The plots on the right reveal a very non-trivial
dependence of the mass corrections from the radial distance, and
overall the impact of these corrections is much larger than the NLO
corrections in the effective theory.

\begin{figure}[t!]
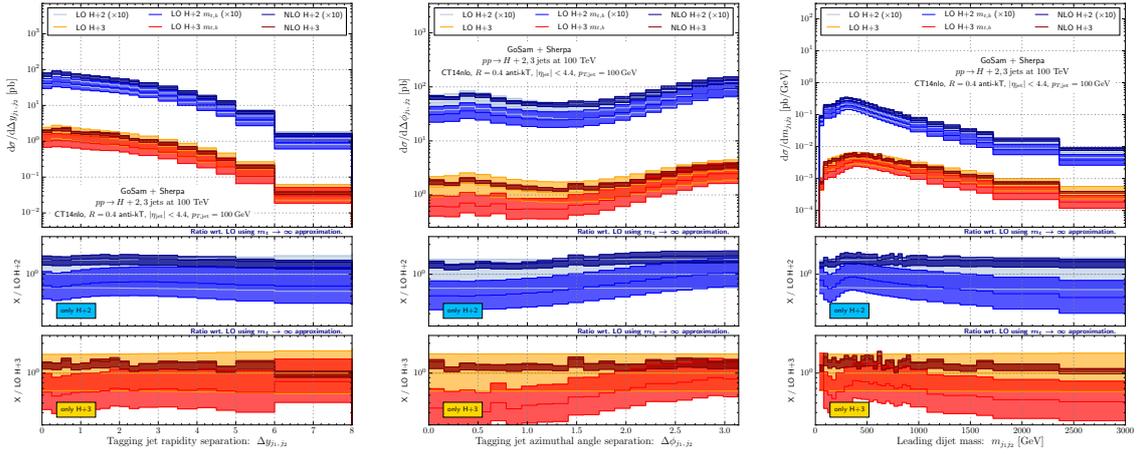

  \centering
  \includegraphics[width=0.327\textwidth]{./figs/%
    Multimeff_H+23_pT100eta44mult_jet_jet_dy_12_40-E100.pdf}
  \hfill
  \includegraphics[width=0.327\textwidth]{./figs/%
    Multimeff_H+23_pT100eta44mult_jet_jet_dphi_12_20-E100.pdf}
  \hfill
  \includegraphics[width=0.327\textwidth]{./figs/%
    Multimeff_H+23_pT100eta44mult_jet_jet_mass_12_300-E100.pdf}
  \caption{\label{fig:fccyjj}%
    Rapidity and azimuthal angle separation between the leading jets
    as well as the leading dijet mass for the higher jet threshold.}
\end{figure}

As can easily be foreseen, the increased deviation of the full SM
predictions from the effective theory is present in the observables
that we will discuss in the following. We will only present plots for
which $p_{T,\,\mathrm{jet}}>100$~\GeV, where the effects are much more
visible. In Figure~\ref{fig:fccyjj} we present the two components
which combined give rise to the radial distance discussed above: on
the left the rapidity separation $\Delta y_{j_1,j_2}$ and in the
center the azimuthal angle separation $\Delta\phi_{j_1,j_2}$ between
the two leading jets. In the former case the mass corrections are
roughly constant over the full kinematical range, whereas in the
latter case they are much larger for small angle separation, and
become almost negligible when the two jets are back-to-back in
azimuth. The reason is similar to the one outlined previously for
$\Delta R_{j_1,j_2}$. On the right of the same figure we show the
leading invariant dijet mass. As already stressed previously, this
observable is particularly important when studying VBF
scenarios. Compared to the curve shown for 13~\TeV and a transverse
momentum cut of 30~\GeV (Fig.~\ref{fig:mjj}), where the mass
corrections barely affected the distribution, we observe now a clear
decrease of the cross section, which reaches $-50\%$ for invariant
masses of the order of $3$~\TeV.

\begin{figure}[b!]
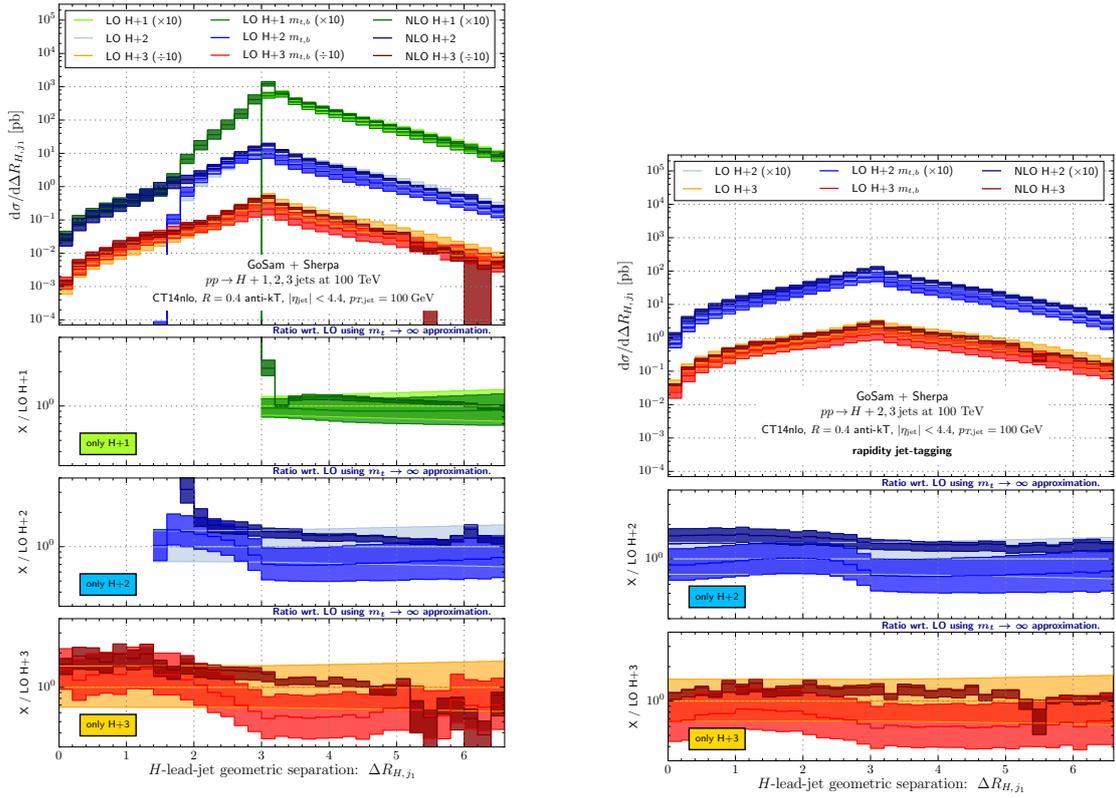

  \centering
  \includegraphics[width=0.47\textwidth]{./figs/%
    Multimeff_H+123_pT100eta44mult_higgs_jet_dR_1_35-E100.pdf}
  \hfill
  \includegraphics[width=0.47\textwidth]{./figs/%
    Multimeff_H+23_pT100eta44mult-tag_higgs_jet_dR_1_35-E100.pdf}
  \caption{\label{fig:fccRhj}%
    Geometric separation between the Higgs boson and the leading
    transverse momentum jet for $p_{T,\,\mathrm{jet}}>100$~\GeV and
    using $p_T$-jet tagging (left) as well as $y$-jet tagging
    (right).}
\end{figure}

Figure~\ref{fig:fccRhj} shows the radial separation between the Higgs
boson and the leading jet for the two different jet tagging
strategies. The plot on the left shows $\Delta R_{H,j_1}$ when using a
$p_T$-jet tagging strategy, whereas on the right we apply the $y$-jet
tagging, which by definition needs the presence of at least two
jets. This observable demonstrates that mass effects can lead to
fairly complicated corrections with respect to the effective theory
predictions. Apart from the \Hj predictions, which because of the
presence of only a single jet are not affected too much by mass
corrections, the full SM predictions increase the cross section at
small radial distance and decrease it for larger values of $\Delta
R_{H,j_1}$. The differences are slightly milder in the right plot when
considering a $y$-jet tagging strategy. This is due to the fact that
the tagging jets in the rapidity tagging are not necessarily hard
jets, which means that the phase space region of hard jets is rather
diluted across the observable.

\begin{figure}[t!]
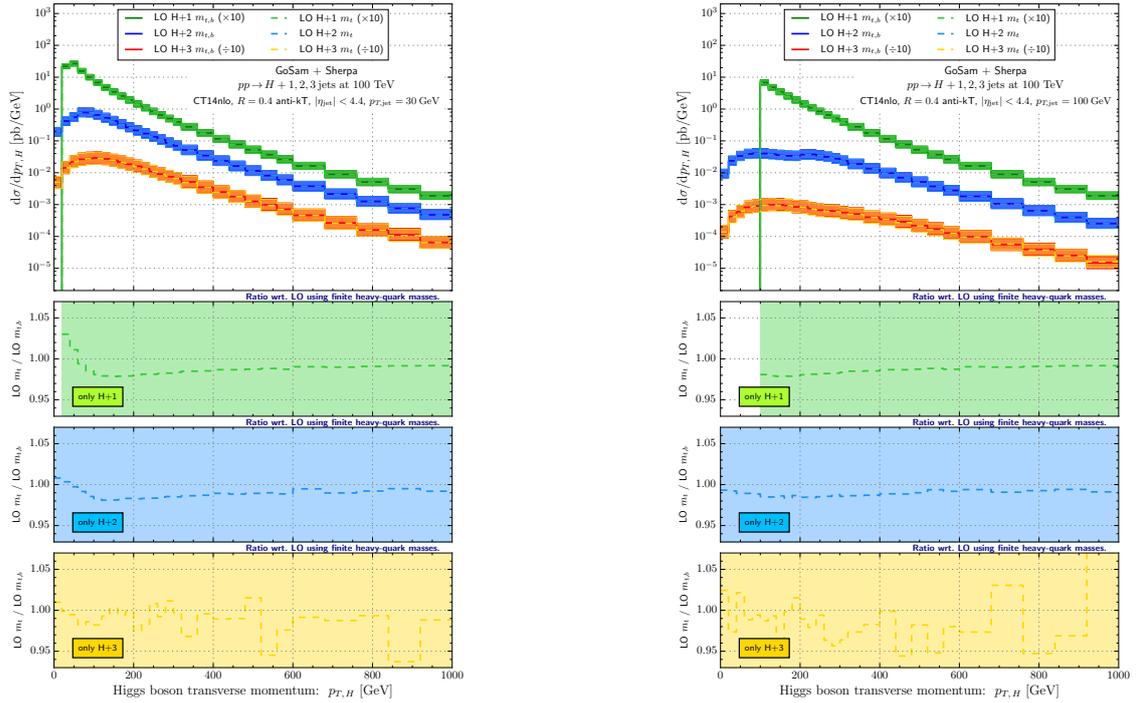

  \centering
  \includegraphics[width=0.42\textwidth]{./figs/%
    Mtbmeff_H+123_pT30eta44mult_higgs_pt_100vbf-E100.pdf}
  \hfill
  \includegraphics[width=0.42\textwidth]{./figs/%
    Mtbmeff_H+123_pT100eta44mult_higgs_pt_100vbf-E100.pdf}
  \caption{\label{fig:fccmtbvsmt}%
    Same as Fig.~\ref{fig:mtb_mt_pts}-left but for a center-of-mass
    energy of 100~\TeV and for $p_{T,\,\mathrm{jet}}>30$~\GeV (left)
    and $p_{T,\,\mathrm{jet}}>100$~\GeV (right).
 }
\end{figure}

To conclude we compare the Higgs boson transverse momentum using full
SM predictions with and without bottom-quark
loops. Figure~\ref{fig:fccmtbvsmt} shows the two results for a minimum
jet transverse momentum of $p_{T,\,\mathrm{jet}}>30$~\GeV on the left,
and for $p_{T,\,\mathrm{jet}}>100$~\GeV on the right. The ratios allow
to appreciate the difference between the two predictions, which is
relevant mainly for \Hj and when the looser jet cut is
used. Increasing the jet cut or the final state multiplicity leads to
a flatter ratio in which the predictions with only top-quark loops are
lower than the one with top- and bottom-quark by $0.5-2\%$.\\

\section{Conclusions}\label{sec:conclusions}

The production of a Higgs boson in association with jets in gluon
fusion is one of the key processes in precision Higgs
physics. Accurate theoretical predictions are fundamental for a
detailed understanding of the electroweak symmetry breaking mechanism.
Usually calculations of higher order corrections to the production of
a Higgs boson in association with jets rely on the approximation of an
infinitely heavy top quark. In this paper we have computed the cross
section at LO in perturbation theory in the full Standard Model
considering a Higgs boson coupling to both top-quark and bottom-quark
loops, including the interference between the two. Furthermore we have
compared these results to the NLO predictions in the effective theory
approach. We give quantitative predictions for a variety of
observables for \Hj, \Hjj and \Hjjj, confirming that
transverse-momentum related observables are particularly affected by
these corrections for values above the top mass. We have calculated
predictions for two center-of-mass energies, for the LHC at $13$~\TeV
and for a possible future circular collider of $100$~\TeV. For the
LHC, we also investigated the impact of finite mass effects when VBF
selections cuts are applied on the tagging jets, in order to enhance
the VBF signal. We find that mass effects typically play an important
role leading to deviations up to one order of magnitude. The breakdown
of the effective theory predictions is driven by the particle with the
highest transverse momentum in the event and is largely independent of
the final state multiplicity. This is of course highly dependent on
the specific observable and scenario under consideration. In
particular, since the corrections affect the harder transverse
momentum regions, choosing a harder $p_T$-cut in the analysis results
in larger mass effects for all observables. We further find that the
effect of including massive bottom-quarks in the loop has a mild
impact. For the total cross section, the bottom-quark contribution
(including its interference with top-quark loops) is around one
percent for a $13$~\TeV LHC. Applying VBF cuts does not lead to
significant changes, which is also true for pp-collisions at a
$100$~\TeV collider. The bottom-quark effects are particularly visible
in the low energy region, where they can lead to deviations of up to
five percent.

In summary, the inclusion of mass effects and their control at an
accuracy beyond leading order will be indispensable for reliable
predictions for both the LHC, but even more for a future collider with
considerably higher center-of-mass energy.

\section*{Acknowledgments}
We wish to thank Gudrun Heinrich, Joey Huston, Michelangelo Mangano
and Simone Marzani for fruitful discussions and comments regarding the
manuscript. GL and JW thank the Max Planck Institute for Physics in
Munich for great hospitality while parts of this work were completed.
NG and JW are also thankful to the Theory Department at CERN and
acknowledge funding by the LPCC. The numerical computations were
performed at the RZG -- the Rechenzentrum Garching near Munich. The
work of NG and MS was supported by the Swiss National Science
Foundation (SNF) under Contract No.~PZ00P2--154829 and No.~PP00P2--128552,
respectively. The work of SH was supported by the US Department of
Energy under Contract No.~DE--AC02--76SF00515. JW acknowledges support
by the National Science Foundation under Grant No.~PHY--1417326.

\appendix
\section{\textsc{Root} Ntuples}
\label{app:ntuples}

In this section we briefly describe an extension of the Ntuples
format, which was introduced recently and partially used for the study
presented here.

The original information stored in the Ntuples entries is summarized
in Table~\ref{tab:ntuple_format}. Recently it was however realized,
that a small upgrade could lead to a broader range of applicability
and to more flexibility. This was mainly driven by the wish of being
able to apply a MiNLO-type~\cite{Hamilton:2012np,Badger:2016bpw} scale
setting on the events stored in the Ntuples. Furthermore, it is of
advantage to have the possibility to change a posteriori the matrix
element weight stored in the Ntuples, while keeping the same set of
events.

\begin{center}
  \begin{table}[bth!]
    \centering
    \begin{tabular}{@{}l*{15}{l}}
    \hline
    \hline
    Branch                  & Description \\
    \hline
    \textit{id}             & Event ID to identify correlated real sub-events                             \\
    \textit{nparticle}      & Number of outgoing partons                                                  \\
    \textit{E/px/py/pz}     & Momentum components of the partons                                          \\
    \textit{kf}             & Parton PDG code                                                             \\
    \textit{weight}         & Event weight, if sub-event is treated independently                         \\
    \textit{weight2}        & Event weight, if correlated sub-events are treated as single event          \\
    \textit{me\_wgt}        & ME weight (w/o PDF), corresponds to ’weight’                                \\
    \textit{me\_wgt2}       & ME weight (w/o PDF), corresponds to ’weight2’                               \\
    \textit{id1}            & PDG code of incoming parton 1                                               \\
    \textit{id2}            & PDG code of incoming parton 2                                               \\
    \textit{fac\_scale}     & Factorisation scale                                                         \\
    \textit{ren\_scale}     & Renormalisation scale                                                       \\
    \textit{x1}             & Bjorken-x of incoming parton 1                                              \\
    \textit{x2}             & Bjorken-x of incoming parton 2                                              \\
    \textit{x1p}            & x’ for I-piece of incoming parton 1                                         \\
    \textit{x2p}            & x’ for I-piece of incoming parton 2                                         \\
    \textit{nuwgt}          & Number of additional ME weights for loops and integrated subtraction terms  \\
    \textit{usr\_wgt[nuwgt]}& Additional ME weights for loops and integrated subtraction terms            \\
    \hline
    \hline
    \end{tabular}
    \caption{\label{tab:ntuple_format}%
      Branches format of the Ntuples files as generated by
      \textsc{Sherpa}.}
  \end{table}
\end{center}

These two requirements led to the development of a new Ntuple format
which we will refer to as \textit{EDNtuples} (Exact Double
Ntuples). In this new format an entry called \textit{ncount}, was
introduced, which keeps track of the number of trials between two good
events during generation. This allows for an exact statistical
treatment when events are reprocessed a posteriori. Furthermore the
momenta are stored in double precision instead of float, to allow for
a more precise kinematical reconstruction. This is needed for example
when the branching history of an event is reconstructed for the MiNLO
scale choice. Furthermore, in order to correctly map the subtraction
counter-events to the appropriate real radiation events when
performing the clustering in MiNLO, additional branches to store the
information on the initial state flavours in the subtraction events
had to be created. Finally, to be able to change the matrix element
weight a posteriori, the phase space weight, which is already
multiplied with the weight coming from the amplitude in the branches
called \textit{me\_wgt} and \textit{me\_wgt2}, is now also stored
separately, giving the possibility to change the weight of the
amplitude. This last extension was of particular interests for this
work since the events stored in the Ntuples could be reused when
computing the amplitude in the full Standard Model theory, when the
heavy quark loops are present at LO accuracy. The new entries added in
the EDNtuples are summarized in Table~\ref{tab:edntuple_format}.

\begin{center}
  \begin{table}[th!]
    \centering
    \begin{tabular}{@{}l*{15}{l}}
      \hline
      \hline
      Branch                  & Description \\
      \hline
      \textit{ncount}         & Number of trials between the previous and current event during generation   \\
      \textit{ps\_wgt}        & Phase space weight                                                          \\
      \textit{id1p}           & PDG code of incoming parton 1 in subtraction event                          \\
      \textit{id2p}           & PDG code of incoming parton 2 in subtraction event                          \\
      \hline
      \hline
    \end{tabular}
    \caption{\label{tab:edntuple_format}%
      Additional new branches introduced for the EDNtuples.}
  \end{table}
\end{center}

\bibliographystyle{JHEP}
\providecommand{\href}[2]{#2}\begingroup\raggedright\endgroup
\end{document}